\documentclass[fleqn,usenatbib]{mnras}

\usepackage{newtxtext,newtxmath}
\usepackage{xcolor}
\usepackage[T1]{fontenc}
\usepackage{physics}
\newcommand{\comment}[1]{}        
\DeclareRobustCommand{\VAN}[3]{#2}
\let\VANthebibliography\thebibliography
\def\thebibliography{\DeclareRobustCommand{\VAN}[3]{##3}\VANthebibliography}

\usepackage{graphicx}	
\usepackage{amsmath}
\newcommand{\bq}{\boldsymbol{q}}     
\newcommand{\bk}{\boldsymbol{k}}   

\title[Interacting dark energy from the joint analysis]{Interacting dark energy from the joint analysis of the power spectrum and bispectrum multipoles with the EFTofLSS}

\author[Maria Tsedrik et al.]{
Maria Tsedrik,$^{1}$\thanks{E-mail:  mtsedrik@ed.ac.uk}
Chiara Moretti,$^{1,2,5}$
Pedro Carrilho,$^{1}$
Federico Rizzo,$^{2,3,4}$
Alkistis Pourtsidou$^{1,5}$
\\
$^{1}$Institute for Astronomy, The University of Edinburgh, Royal Observatory, Edinburgh EH9 3HJ, UK\\
$^{2}$Istituto Nazionale di Astrofisica, Osservatorio Astronomico di Trieste, via Tiepolo 11, 34143 Trieste, Italy \\
$^{3}$Istituto Nazionale di Fisica Nucleare, Sezione di Trieste,  via  Valerio  2,  34127 Trieste,  Italy \\
$^{4}$Institute for Fundamental Physics of the Universe, Via Beirut 2, 34151 Trieste, Italy\\
$^{5}$Higgs Centre for Theoretical Physics, School of Physics and Astronomy,
The University of Edinburgh, Edinburgh EH9 3FD, UK 
}

\date{Accepted XXX. Received YYY; in original form ZZZ}

\pubyear{2022}

\begin{document}
\label{firstpage}
\pagerange{\pageref{firstpage}--\pageref{lastpage}}
\maketitle

\begin{abstract}
Interacting dark energy models have been suggested as alternatives to the standard cosmological model, $\Lambda$CDM. 
We focus on a phenomenologically interesting class of dark scattering models that is characterised by pure momentum exchange between dark energy and dark matter. 
This model extends the parameter space with respect to $\Lambda$CDM by two parameters, $w$ and $A$, which define the dark energy equation of state and the strength of the coupling between dark energy and dark matter, respectively. 
In order to test non-standard cosmologies with Stage-IV galaxy clustering surveys, it is crucial to model mildly nonlinear scales and perform precision vs accuracy tests. We use the Effective Field Theory of Large-Scale Structure, and we perform validation tests by means of an MCMC analysis using a large set of N-body simulations. We find that adding the bispectrum monopole to the power spectrum multipoles improves the constraints on the dark energy parameters by $\sim 30 \%$ for $k_{\mathrm{max}, B}^{l=0} = 0.11$ $h$ Mpc$^{-1}$, without introducing biases in the parameter estimation.  We also find that the same improvement can be achieved with more moderate scale cuts and the use of bias relations, or with the addition of the bispectrum quadrupole. Finally, we study degeneracies between the dark energy parameters and the scalar amplitude $A_\mathrm{s}$ and discuss the corresponding projection effects, as well as degeneracies with other cosmological parameters.

\end{abstract}


\section{Introduction}
\label{sec:intro}

In the next few years, Stage-IV galaxy surveys 
such as the Dark Energy Survey (DES)\footnote{\url{https://www.darkenergysurvey.org/}}~\citep{Abbott:2021bzy}, the Dark Energy Spectroscopic Instrument (DESI)\footnote{\url{https://www.desi.lbl.gov/}}~\citep{Aghamousa:2016zmz}, 
Euclid\footnote{\url{http://euclid-ec.org}}~\citep{Laureijs:2011gra}, the Nancy Grace Roman Space Telescope\footnote{\url{https://www.nasa.gov/roman}}~\citep{spergel2015wide}, and the Vera C. Rubin Observatory’s Legacy Survey of Space and Time (LSST)\footnote{\url{https://www.lsst.org/}}~\citep{LSST:2008ijt}
promise to provide high precision measurements, which will allow us to study the large-scale structure of the Universe with unprecedented accuracy. A key goal of these surveys is to test for deviations from the standard cosmological model, $\Lambda$CDM. The $\Lambda$CDM model postulates two exotic components: cold dark matter (CDM), and dark energy in the form of a cosmological constant ($\Lambda$). The nature of these dark sector components is unknown. Moreover, in recent years issues related to tensions in the determination of the $H_0$ and $\sigma_8$ cosmological parameters between early and late Universe probes have emerged. The search for alternatives to $\Lambda$CDM that could also explain the apparent tensions has led to a plethora of exotic dark energy, dark matter, and modified gravity theories \citep[for reviews see][]{ CopelandDarkEnergy, BertonDarkMatter, MGReview, TensionsReview}. In preparation for the stringent tests of these theories with forthcoming data, validation tests of the exotic models against simulations have to be performed. One of these non-standard cosmological models, called Interacting Dark Energy, is the focus of this study.

The term Interacting Dark Energy (IDE) describes a broad class of models, in which a non-gravitational coupling between dark energy and dark matter is allowed. A large suite of such models can be constructed by making different choices for the form of the coupling function \citep[for instance,][]{CoupledQuintessence, Farrar2004, Clemson:2011an, AlkistisIDE2013}. Interacting dark energy models have also been considered as candidate models to alleviate the $\sigma_8$ and $H_0$ tension, with various degrees of success \citep{AlkistisTram2016, IDESimulations2017, CoupledQuintSigma8Tension, IDEH0sigma8Tension, IDEH0tension, IDEsigma8Tension, IDETensionsLuca, IDES8}. 
The vast majority of models exhibits background energy transfer. As a result, they are severely constrained by Cosmic Microwave Background (CMB), Baryonic Acoustic Oscillations (BAO) and Supernovae (SNeIa) measurements \citep[for instance,][]{2009PhRvD..80j3514X, IDEconstr3, IDEconstr4, Pan:2019gop, IDEconstr2, IDEconstr1, Nunes:2022bhn}. In this work we consider a popular class of IDE models, which is uncoupled in the background and exhibits only momentum exchange between dark energy and dark matter, at the level of the linear perturbations. This feature makes the model capable of fitting both large-scale structure (LSS) and CMB measurements extremely well for a wide range of couplings \citep{AlkistisTram2016, IDESimulations2017, AlkistisAlessio2021}.

In contrast to modified gravity (MG) theories, IDE models do not influence the speed of gravitational waves or the gravitational potential of planets, hence they do not need screening mechanisms in order to surpass Solar System tests. Furthermore, instead of an enhancement in the growth of structure at nonlinear scales \citep[as in the case of popular MG models, e.g.,][]{fR_Review}, some interacting models predict a suppression of growth and hence lead to an alleviation of the $\sigma_8$ tension.

Non-standard cosmologies generally have a larger parameter space (more free parameters) than $\Lambda$CDM, and these may result to observable effects at linear and nonlinear scales. 
For example, \cite{IDESimulations2015} showed that nonlinear scales are much more informative in terms of constraints for dark matter - dark energy scattering than the linear regime. This highlights the importance of accurately modelling nonlinear scales for extended models. The specific IDE model under consideration in this work was studied by \cite{PedroCola}. The authors used a set of approximate (COLA) simulations run with standard cosmology to perform null tests of three perturbative approaches to compute the halo power spectrum multipoles. The authors conclude that a) a false $\sim 3\sigma$ detection of exotic dark energy would occur should the nonlinear modelling be incorrect b) precise and accurate constraints on IDE parameters are possible when appropriate scale cuts are considered, and c) the Effective Field Theory of Large-Scale Structure (EFTofLSS) approach \citep{EFTBaumann, EFTSenatore, GuidoBOSS, MinervaBispectrumRSD} provides a model accurate up to a wave-number of $k_\mathrm{max} \sim 0.3$ $h$ Mpc$^{-1}$. In this study we apply the same model and extend upon the previous work by adding the bispectrum monopole and quadrupole to the analysis, and by using better simulations.

The joint analysis of the power spectrum and bispectrum has been considered for Stage-IV surveys analyses that aim to constrain $\Lambda$CDM. The addition of the bispectrum is useful for breaking bias degeneracies and tightly constraining bias parameters. An example is the significant increase in the constraining power on the mass fluctuation amplitude $A_\mathrm{s}$ when the real space analysis includes the bispectrum \citep{IntegrationBinning}: indeed, the bispectrum is able to break the strong degeneracy between $A_\mathrm{s}$ and the linear bias, $b_1$.
This is however no longer the case in redshift space, since the aforementioned degeneracy is already broken at the level of the power spectrum multipoles. There are various examples of the joint analysis in redshift space in the literature. To name a few, a Fisher Matrix forecast analysis was performed in \cite{Vica} with tree-level Standard Perturbation Theory (SPT) \citep{SPT}; a full likelihood analysis was done in \cite{BispAniso1} and \cite{BispAniso2} with the TNS modelling \citep{TNS} and the Quijote suite of N-body simulations \citep{QuijoteSim}, as well as in \citet{GuidoBOSS,PhilcoxBOSS,EFTIvanovBispectrum}, with data from the BOSS survey \citep{BOSSData} and the EFTofLSS model. 
The above forecasts and data analyses concentrated on the $\Lambda$CDM model.
In this paper we analyse the informational content of the bispectrum in addition to the power spectrum in redshift space for a non-minimal, interacting dark energy model.
To our knowledge, this is the first analysis of this kind for exotic dark energy.

This paper is organized as follows: \autoref{sec:DE_model} is dedicated to the specific interacting dark energy model we use in our work, its main features and impact on structure formation; in \autoref{sec:PB_model} we describe the EFTofLSS model for the power spectrum and bispectrum with a brief overview on the nuisance parameters and their co-dependencies. In \autoref{sec:MCMC} we describe our likelihood pipeline and performance metrics. Our results are presented in \autoref{sec:results}. We conclude in \autoref{sec:conclusion}.

\section{Interacting Dark Energy Modelling}
\label{sec:DE_model}

In this work we focus on the subclass of IDE models which exhibits no coupling at the background level and introduces a pure momentum transfer at the level of linear perturbations. In general, non-gravitational interaction between dark matter and dark energy can be introduced either at the level of the action \citep{AlkistisIDE2013, Tamanini:2015iia,  IDEsigma8Tension} or more phenomenologically at the level of the fluid equations \citep[e.g.,][]{Valiviita2008, Simpson2010}. This leads to two different approaches for our model of interest.

The first approach corresponds to the ``Type 3'' class of models from the action-based derivation in \cite{AlkistisIDE2013}. There, the pure exchange of momentum is generated by a coupling between the dark matter velocity field and the covariant derivative of the dark energy scalar field. In this case, the two parameters describing the extension with respect to the standard cosmological model are given by the coupling strength $\beta$ and the exponential factor of the quintessence potential $\lambda$. These parameters  are included in the action as follows:
\begin{equation}
    S_\phi = \int \dd t ~ \dd^3 x ~ a^3 \left[ \frac{1}{2}(1 - 2\beta) \dot{\phi}^2 - \frac{1}{2} |\nabla \phi|^2 - V_0 \mathrm{e}^{-\lambda \phi}  \right]\, ,
\end{equation}
where $\phi$ is the scalar field (quintessence), the last term is the quintessence potential of the single exponential form, the dot represents a derivative with respect to cosmic time and $a$ is the scale factor. This model is explicitly studied in \cite{AlkistisTram2016}, \cite{AlkistisAlessio2021} and has been implemented into the CLASS \citep{CLASS, Blas:2011rf} and CAMB \citep{CAMB} Boltzmann codes. These analyses vary $\beta$ and $\lambda$, while the quintessence potential normalisation, $V_0$, is varied automatically by the Boltzmann solver in order to match a given $\Omega_\phi$ today. 

The second approach is derived in the elastic scattering formalism by \citet{Simpson2010}. The main idea is that a dark matter particle moves through the isotropic dark energy fluid and experiences a drag-force proportional to the scattering cross-section between the particle and the fluid. Since a dark matter particle is more massive than the energy exchange with the dark energy fluid, they interact via elastic scattering, in a way that resembles Thomson scattering between non-relativistic electrons and photons in the radiation-dominated era.

Such interaction involves no energy transfer and hence it leaves the energy conservation equation unchanged with respect to the standard cosmology. 
Moreover, we assume that the dark energy density and velocity fields are homogeneous (i.e., $\delta_\mathrm{DE} = \theta_\mathrm{DE} = 0$), since dark energy perturbations are damped within the cosmic horizon. This follows from the assumption of the dark energy speed of sound being equal to the speed of light, and was confirmed numerically in \citet{IDESimulations2015}. While these fluctuations become non-negligible at super-horizon scales, even the most extreme modifications do not contradict the CMB measurements due to the presence of cosmic variance \citep{AlkistisTram2016}.    
In presence of momentum exchange between the two components of the dark sector, and with the approximation of homogeneity in dark energy, this IDE model modifies only the Euler equation to
\begin{equation}
\label{eq:Euler}
    a~\partial_a \Theta + \left( 2 + (1+w)~\xi~ \frac{\rho_\mathrm{DE}}{H} + \frac{a~\partial_a H}{H} \right) \Theta + \frac{\nabla^2 \Phi}{a^2 H^2} = 0 \, ,
\end{equation}
where $\xi = \sigma_\mathrm{D}/m_\mathrm{c}$ with $\sigma_\mathrm{D}$ is the cross-section of the dark energy - dark matter interaction and $m_\mathrm{c}$ the mass of dark matter particle, and $\Theta = \theta_\mathrm{c} /a H$ with $\theta_\mathrm{c}$ the dark matter velocity divergence. $H = \dot{a}/a$ is the Hubble rate and it describes the expansion history in the flat universe composed only of dark matter and dark energy by
\begin{equation}
\label{eq:H}
    H^2 = H_0^2 \left( \Omega_{\mathrm{c}, 0}~ a^{-3} + \Omega_{\mathrm{DE}, 0}~ a^{-3(1+w)} \right)  \, ,
\end{equation}
where as above $w$ is the dark energy equation of state parameter. $H_0$, $\Omega_{\mathrm{c}, 0}$, $\Omega_{\rm{DE}, 0}$ are the values of Hubble parameter, dark matter density and dark energy density parameters today. Additionally, the gravitational potential stays unchanged and is given by the following Poisson equation:
\begin{equation}
    \nabla^2 \Phi = \frac{3}{2} a^2 H^2 \Omega_\mathrm{c} \delta_\mathrm{c} \, ,
\end{equation}
where $\Omega_\mathrm{c}$ is the background dark matter density parameter at the scale factor $a$ and $\delta_\mathrm{c}$ is the corresponding density contrast.

\begin{figure}
\centering
	\includegraphics[width=0.85\linewidth]{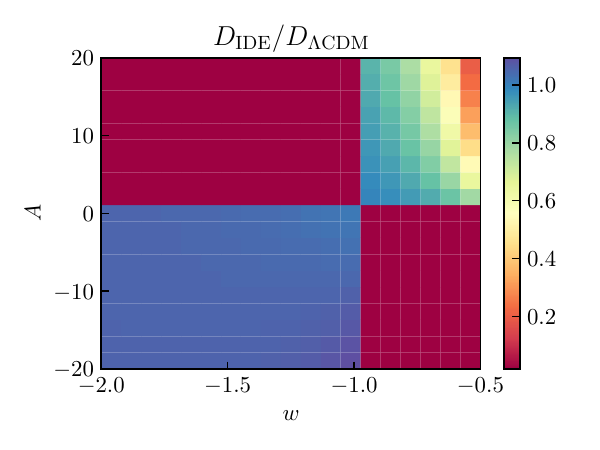}
	\includegraphics[width=0.85\linewidth]{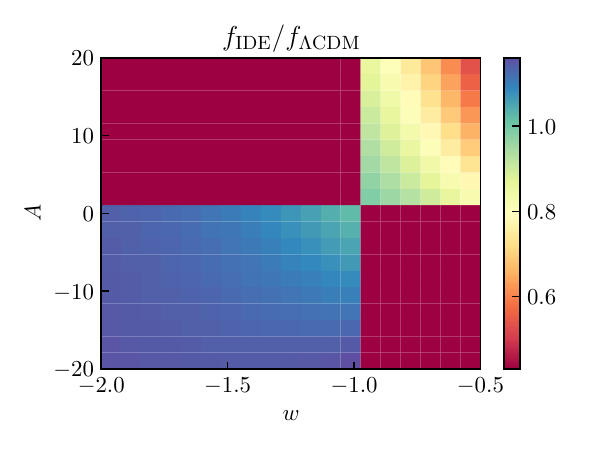}
    \caption{\textit{Top panel}: the linear growth factor fractional deviation relative to the $\Lambda$CDM case with $D_{\Lambda\mathrm{CDM}} = 0.475$ for varying values of IDE parameters, $w$ and $A$. \textit{Bottom panel}: the logarithmic growth rate fractional deviation from the $\Lambda$CDM value $f_{\Lambda\mathrm{CDM}} = 0.861$. Both growth parameters are computed at redshift $z=1$. Parameter $A$, defining the coupling strength between dark energy and dark matter, is given in the units of b GeV$^{-1}$. The red patches in the upper left and the lower right corners denote the forbidden area of the IDE parameter space due to the condition $A/(1+w)\geq 0$.}
    \label{fig:IDEfD}
\end{figure}
Although there is no direct mapping between the elastic scattering and the ``Type 3'' models, a particular subclass of ``Type 3'' models can reproduce the drag-like behaviour \citep{Skordis2015}. In \citet{IDESimulations2017} the authors show how these models can be matched approximately. In this work we use the phenomenological approach with parameters $w$ and $A = \xi (1+w)$, which characterise the equation of state for dark energy and the coupling strength, respectively. We assume that $w$ is constant, it affects the expansion history in \autoref{eq:H} as well as structure formation in \autoref{eq:Euler}. At the same time, $A$ impacts only the perturbation growth and is chosen in this form (i.e., a combination of $\xi$ and $w$) for the purposes of parameter inference. This is particularly important for $w \approx -1$, where we can constrain $A$ while $\xi$ could take an arbitrary large value.  

Parameters $w$ and $A$ modify the linear growth factor $D(a)$ and the logarithmic growth rate $f = \dd \ln{D}/ \dd \ln{a}$. For a constant value of $w$ in the late Universe with redshift $z \leq 10$, in the canonical (phantom) case of $w_\mathrm{DE} > -1 \, (w_\mathrm{DE} < -1)$ the growth factor is decreased (increased) with respect to $\Lambda$CDM. The effect gets more prominent as the redshift decreases, and for larger values of $\xi$. This leads to a scale-independent suppression (enhancement) of the power spectrum at large scales. On smaller scales, the modification comes from the drag term in \autoref{eq:Euler}: for $w_\mathrm{DE} > -1 \, (w_\mathrm{DE} < -1)$ there is a dissipation (injection) of kinetic energy within collapsed structures. In other words, for $w_\mathrm{DE} > -1 \, (w_\mathrm{DE} < -1)$ particles inside collapsed objects experience additional friction (drag), hence it becomes harder (easier) for them to stream away, with a strong and scale-dependent enhancement (suppression) of the growth of structure in the nonlinear regime. This behaviour, anticipated from the theory, has been shown to hold true in N-body simulations by \citet{IDESimulations2015} and \citet{IDESimulations2017}.

In this work we consider only one redshift $z=1$, at which the impact of the IDE parameters is observable but not as prominent as at lower redshifts. In \autoref{fig:IDEfD} we show the impact of different values of $w$ and $A$ on the growth parameters at the redshift of our interest. The ratios are taken with respect to the $\Lambda$CDM values: $D_{\Lambda\mathrm{CDM}} = 0.475$ and $f_{\Lambda\mathrm{CDM}} = 0.861$ for our choice of cosmological parameters. The red patches in the upper left and the lower right corners denote the forbidden area of the IDE parameter space: the sign of $A$ and $(1+w)$ has to be the same, since $\xi = A/(1+w)$ from \autoref{eq:Euler} has to be non-negative. For $w<-1$ dark energy starts to dominate over matter at later times than in $\Lambda$CDM, hence with decreasing $w$ the growth rate quickly becomes $1$ (as in the matter-dominated epoch) nearly independent from the value of $A$. The linear growth factor is also independent of $A$ and is slightly larger than the $\Lambda$CDM case for $w<-1$. For $w>-1$ the expected power suppression at linear scales is present, with a stronger effect for larger values of $A$.

\section{Power Spectrum and Bispectrum Modelling}
\label{sec:PB_model}

Halos and galaxies are biased tracers of the matter density field \citep{BiasReview}. Their clustering properties up to mildly nonlinear scales can be described using the perturbative framework. The EFTofLSS formalism is based on perturbation theory, with the addition of counterterms to account for the impact of unknown small-scale physics, such as galaxy formation, on large-scale modes. The functional form of the counterterms is specified by the symmetries of the density field, while the nuisance parameters are determined by the data. We note that application to data would not be possible without huge effort in the development on the theoretical side (see for example \citealt{SPT, Crocce:2007dt, TNS, EFT1,EFT2, Vlah:2015sea, EFT3,EFT4, Vlah:2018ygt}, and for a more comprehensive list for the development of the EFTofLSS see the footnote in \citealt{EFTBispectrumOneLoop}).

\subsection{Bias expansion}
\label{subsec:Bias}

Both dark matter halos (the focus of this study) and galaxies are biased tracers of the underlying matter distribution. The halo overdensity can be expanded as $\delta_h = \sum_n b_n /n! \delta^n $, where $\delta^n$ are higher powers of the matter density contrast and $b_n$ are the corresponding local bias terms \citep{BiasLocalExp}. Halos are formed by the gravitational collapse of matter from a spatially finite region. Anisotropies created by this process impact the local distribution of objects, and thus the bias parameters, with tidal effects \citep{TidalBias1, TidalBias2} and introduce scale dependency by breaking the local assumption \citep{BiasMcDonald}. Effects from an ellipsoidal gravitational collapse are encoded in the non-local operators, dubbed $\mathcal{G}_2$ and $\Gamma_3$: the first one represents the tidal stress tensor produced by the gravitational potential, while the second one denotes the difference between the tidal stress tensors from the gravitational and velocity potentials. Effects due to the finite size of the collapsing region are taken into account by the higher-order derivative of the matter field $\nabla^2 \delta$. Additionally, the relation between matter and galaxies is also affected by stochasticity, whose impact can be modeled as an additional contribution to Poisson shot noise on large scales \citep{SubPoisson1, SubPoisson2}, while on small scales it can exhibit a scale-dependence due to the halo-exclusion effect \citep{HaloStohastic,TestinBiasNkNoise_bG2rel}. We consider the bias expansion given by \citep{BiasReview, RenormBias}:
\begin{equation}
\label{eq:biases}
    \delta_h = b_1 \delta + \frac{b_2}{2} \delta^2 +b_{\mathcal{G}_2} \mathcal{G}_2 + b_{\Gamma_3} \Gamma_3 + b_{\nabla^2 \delta } \nabla^2 \delta + \epsilon + \epsilon_\delta \delta \, ,
\end{equation}
where $b_1$ and $b_2$ are the linear and quadratic bias, and the last two terms correspond to stochasticity contributions. Note that the higher derivative term gives rise to a $k^2 P_L(k)$ correction, which is degenerate with one of the counterterms. Therefore, we omit $b_{\nabla^2 \delta}$ in our analysis, as previously done in \citet{MinervaPSBReal}. Moreover, we omit from \autoref{eq:biases} all operators that do not contribute to our chosen theoretical models for the power spectrum and bispectrum. Note that this bias expansion was derived in the context of $\Lambda$CDM cosmology. However, it is still applicable in our model, since IDE features a scale-independent linear growth, implies the same structure of the SPT kernels (see the next subsection) and obeys the same symmetries as in $\Lambda$CDM \citep[for more information on bias expansion from symmetries see][]{Fujita:2020,DAmico:2021}.

\subsection{Power spectrum and bispectrum}

The mildly nonlinear power spectrum and bispectrum in real space using the EFTofLSS can be modeled with three ingredients: a) the leading-order contribution in SPT at one-loop for the power spectrum  with the addition of the EFTofLSS counterterms, and at tree-level for the bispectrum, b) a bias expansion, and c) an infrared-resummation routine to take the damping of the oscillatory features into account \citep{IRResum1, IRResum2}. 
In addition to that, to model redshift space quantities we should take the distortions due to peculiar velocities into consideration. This introduces an additional dependency of the observables on the direction of separation between the objects with respect to the line-of-sight, $\boldsymbol{\hat{s}}$. For the power spectrum this dependency is characterized by the cosine of the angle formed by the separation wave-number with the line-of-sight, $\mu = (\bk \cdot \boldsymbol{\hat{s}})/k$. For the bispectrum we need two cosines, $\mu_1$ and $\mu_2$, of $\bk_1$ and $\bk_2$ with the line of sight $\boldsymbol{\hat{s}}$, with the condition $\bk_1 + \bk_2 + \bk_3 = \boldsymbol{0}$. However, it is more convenient to describe triangle configurations with three wave-numbers ($k_1$, $k_2$, $k_3$), and two angles, $\theta = \arccos{\mu_1}$ and $\xi$ being the azimuthal rotational angle of $\bk_2$ around $\bk_1$ \citep{ScoccimarroNotation}. In total, the redshift space bispectrum depends on five variables, three of which determine the shape of the triangle while the remaining two define its orientation with respect to the line-of-sight. 

The complete set of equations we use in this work will be detailed in a forthcoming paper by \citet{Chiara}, where we also give a detailed description of the code used to perform the analysis. Here we briefly highlight the main formulae and techniques used. These formulae are the same as in $\Lambda$CDM, as the interaction under consideration only modifies the linear growth factor, $D$ (appearing below via $P_L$) and the linear growth rate of structure, $f$.

The expression for the halo power spectrum in redshift space that we adopt in our analysis is given by
\begin{flalign}
\label{eq:Pgg}
    P_{hh}(k, \mu) =P_\mathrm{SPT}(\bk) + P_\mathrm{ctr}(\bk) + P_\mathrm{stoch} (k) \, ,
\end{flalign}
in which the first term equals 
\begin{flalign}
    P_\mathrm{SPT}(\bk) &= P_\mathrm{Kaiser}(\bk)+P_\mathrm{1-loop}(\bk) = Z_1(\bk) P_L(k)  \nonumber \\ &+ 2 \int \dd^3 \bq [Z_2 (\bq, \bk-\bq)]^2 P_L(q) P_L(|\bk-\bq|) \nonumber\\ 
    &+ 6Z_1(\bk) P_L(k) \int \dd^3 \bq Z_3(\bk, \bq, -\bq) P_L(q) 
\end{flalign}
with $Z_1(\bk)$, $Z_2(\bk_1, \bk_2)$ and $Z_3(\bk_1, \bk_2, \bk_3)$ the redshift space kernels for the SPT loop corrections \citep{ZKernels}, and $P_L$ being the linear power spectrum. Note that the computation is done under the assumption of the Einstein-de Sitter (EdS) approximation, i.e., the SPT kernels are computed with $\Omega_{\mathrm{c}, 0} = 1$ in the absence of dark energy, while the time-dependence of density perturbations is captured by the linear growth factor from the IDE model. The scale-independent linear growth in IDE and the redshift of our interest being $z=1$ allow us to conclude that EdS is a good approximation accurate at the sub-percentage level \citep[see ][for a detailed study of the EdS approximation in redshift space]{Donath:2020abv, Zhang:2022}. 

We then apply IR-resummation \citep{IRResum1, IRResum2}, following the procedure described in detail in \citet{Ivanov:2020}. This gives us the re-summed version of the terms above:
\begin{flalign}
    &P_\mathrm{Kaiser} \rightarrow (b_1+f\mu^2)^2\left[ P_L^\mathrm{nw} + \mathrm{e}^{-k^2 \Sigma^2_\mathrm{tot}}P^\mathrm{w}_L (1+k^2\Sigma^2_\mathrm{tot})\right] \, , \nonumber \\ 
    &P_\mathrm{1-loop} \rightarrow P^\mathrm{nw}_\mathrm{1-loop} + \mathrm{e}^{-k^2 \Sigma^2_\mathrm{tot}}P^\mathrm{w}_\mathrm{1-loop} \, ,
\end{flalign}
where $\Sigma_\mathrm{tot}$ is given by Equation A.6 in \citet{Ivanov:2020}, while the wiggle-no-wiggle splitting is performed using the Einstein-Hu fitting function \citep{EinsteinHu:1998}. The redshift space loop corrections lead to 28 independent integrals, which we compute using the Fast-PT algorithm \citep{FPT1, FPT2}. 

The stochastic power spectrum is given by \citep{TestinBiasNkNoise_bG2rel}
\begin{equation}
\label{eq:Pstoch}
    P_\mathrm{stoch}(k) = (1 + \alpha_P + \epsilon_{k^2} k^2) \bar{n}^{-1}\, ,
\end{equation}
with two free parameters, $\alpha_P$ and $\epsilon_{k^2}$, describing a constant deviation from purely Poisson shot noise $\bar{n}^{-1}$, and higher-order scale dependent corrections generated to account for the short-range non-locality, respectively. The physical meaning of $\alpha_P$ can be explained with the halo exclusion effect \citep{SmithHaloScaleDep, BaldaufHaloStoch}; its value can be either positive (e.g., in galaxy populations with high satellite fractions) or negative (e.g., in central galaxies of massive halos) \citep{BaldaufHaloStoch}. The term $(1+\alpha_P)\, \bar{n}^{-1}$ corresponds to the noise parameter $N$ usually used in the literature, which denotes a scale-independent noise power spectrum. In this work we include scale-dependency into the shot noise, a feature which is supported by numerical simulations \citep{GinzburgNkSims}.
Furthermore, as we show below, the data prefers a strong scale-dependence in the shot noise (see \autoref{subsec:data} and \autoref{fig:data}).

The EFTofLSS counterterms for this model are 
\begin{flalign}
\label{eq:counter}
    P_\mathrm{ctr}(k, \mu) &= -2 \tilde{c}_0 k^2 \tilde{P}_L(k) - 2 \tilde{c}_2 k^2 f \mu^2 \tilde{P}_L(k) \nonumber \\
    &- 2 \tilde{c}_4 k^2 f^2 \mu^4 \tilde{P}_L(k) + P_{\mathrm{ctr}, \nabla^4 \delta} (k, \mu) \, ,
\end{flalign}
with the IR re-summed linear power spectrum
\begin{flalign}
    \tilde{P}_L(k) &= P^\mathrm{nw}_L + \mathrm{e}^{-k^2 \Sigma^2_\mathrm{tot}}P^\mathrm{w}_L
\end{flalign}
and an additional counterterm proportional to $\mu^4k^4\tilde{P}_L(k)$ to include higher-order contributions and model to some extent the Finger of God effect (FoG):
\begin{equation}
    P_{\mathrm{ctr}, \nabla^4 \delta} (k, \mu) = c_{\nabla^4 \delta} k^4 f^4 \mu^4 (b_1 + f \mu^2)^2 \tilde{P}_L(k) \, .
\end{equation}
Overall, this power spectrum model includes in total 10 parameters: \{$b_1$, $b_2$, $b_{\mathcal{G}_2}$, $b_{\Gamma_3}$, $\tilde{c}_0$, $\tilde{c}_2$, $\tilde{c}_4$, $\alpha_P$, $\epsilon_{k^2}$, $c_{\nabla^4 \delta}$ \}.

For the tree-level bispectrum the expression at tree-level is given by
\begin{flalign}
    &B_g(\bk_1, \bk_2, \boldsymbol{\hat{s}}) = B_\mathrm{stoch} (\bk_1, \bk_2, \boldsymbol{\hat{s}}) \nonumber\\
    &+ \sum_{\bk_1 \leq \bk_i \leq \bk_j \leq \bk_3} 2 Z_1(\bk_i) Z_1(\bk_j) Z_2(\bk_i, \bk_j) \tilde{P}_L(k_i) \tilde{P}_L(k_j)  \, ,
\end{flalign}
with the stochastic contribution
\begin{equation}
  B_\mathrm{stoch} (\bk_1, \bk_2, \boldsymbol{\hat{s}}) = (1+\alpha_B) \bar{n}^{-1} \sum_{i = 1}^3 Z_1^2(\bk_i)\tilde{P}_L(k_i) +\bar{n}^{-2} \, ,
\end{equation}
where $\alpha_B$ describes a deviation from the Poisson limit. In contrast to \citet{EFTIvanovBispectrum}, we do not include the FoG modelling and its corresponding parameter in our tree-level model, since such a contribution is expected only at 1-loop level and our data consists of halos only. This leaves the halo bispectrum to be described by 4 parameters: \{$b_1$, $b_2$, $b_{\mathcal{G}_2}$, $\alpha_B$ \}. 

Overall, we express our theoretical prediction in terms of multipoles, in order to eliminate the $\mu$-dependence.
For the power spectrum we compute
\begin{equation}
\label{eq:Pell}
    P_l(k) = \frac{2 l + 1}{2} \int_{-1}^{1} \dd \mu P_{hh}(k, \mu) \mathcal{P}_l(\mu)
\end{equation}
where $\mathcal{P}_l(\mu)$ are the Legendre polynomials of order $l$ and $P_{hh} (k, \mu)$ is the redshift space power spectrum given by  \autoref{eq:Pgg}. In this analysis we use the monopole ($l = 0$), quadrupole ($l = 2$), and hexadecapole ($l = 4$). Studies have shown that these three multipoles contain most of the cosmological information \citep[e.g.,][]{TNS, FlorianMultipoles, Markovic:2019sva}.

An analogous expansion can be applied to the bispectrum, by expanding in terms of spherical harmonics as
\begin{equation}
    B_h( k_1, k_2, k_3, \theta, \xi ) = \sum_l \sum_{m=-l}^l B_{lm} (k_1, k_2, k_3) Y_l^m(\theta, \xi)
\end{equation}
with the same convention for triangle description as above \citep{ScoccimarroNotation}. 
The expansion coefficients are given by
\begin{equation}
    B_{lm}(k_1, k_2, k_3) = \int_{-1}^{1} \dd \cos{\theta} \int_0^{2\pi} \dd \xi B_h( k_1, k_2, k_3, \theta, \xi ) Y_l^m(\theta, \xi) \, .
\end{equation}
As is typical in the literature, when we describe the redshift space multipoles of the bispectrum, we consider only the case with $m=0$, as it has been shown that such seemingly strong compression does not lead to a significant loss of information \citep{BispectrumM0Enough}. This allows us to write the bispectrum version of \autoref{eq:Pell}, since the spherical harmonics reduce to Legendre polynomials in the $m=0$ case:  

\begin{flalign}
    &B_l (k_1, k_2, k_3) = \frac{2 l +1}{2} \nonumber \\
    & \cross \int_{-1}^1 \dd \cos{\theta} \left[ \frac{1}{2\pi} \int_0^{2 \pi} \dd \xi B_h( k_1, k_2, k_3, \theta, \xi ) \right] \mathcal{P}_l(\cos{\theta}) \, .
\end{flalign}

Let us briefly list the known degeneracies, which follow from previous studies, our tests and theory directly:
\begin{enumerate}
    \item $f-b_1$ are anti-correlated: both parameters control the amplitude on large scales, they occur combined with various powers in the model, hence if one gets larger the other should decrease.   
    \item The counterterms $\tilde{c}_0$, $\tilde{c}_2$, $\tilde{c}_4$ and $c_{\nabla^4 \delta}$ are all slightly degenerate with one another, but especially strong is the anti-correlation between $\tilde{c}_4$ and $\tilde{c}_2$. This becomes clear after the projection of \autoref{eq:counter} into multipoles: for the monopole the first three terms in this equation yield $-1/2\,(2\, \tilde{c}_0+0.7\, \tilde{c}_2 f+0.4\, \tilde{c}_4 f^2)\, k^2 P_L$, while for the quadrupole we obtain $-5/2\, (0.3\, \tilde{c}_2 f + 0.2\, \tilde{c}_4 f^2)\, k^2f P_L$. For instance, for $f \approx 1$ the counterterm parameters $\tilde{c}_4$ and $\tilde{c}_2$ cancel each other out if $\tilde{c}_4 \approx - \tilde{c}_2$. In these ad-hoc calculations, we have neglected the angle-dependence in the linear power spectrum due to IR-resummation, which however will affect the counterterms in the same way and does not impact our argument. Additionally,  $\tilde{c}_4$ cannot be constrained without the power spectrum hexadecapole data. This can be seen directly from \autoref{eq:counter} where there are three terms with coefficients $\tilde{c}_0$, $\tilde{c}_2$, $\tilde{c}_4$ that all of them have a form of $k^2 P_L(k)$. From this it follows that we need all three multipoles to constrain all three parameters. 
    \item The nonlinear bias parameters $b_2$, $b_{\mathcal{G}_2}$ and $b_{\Gamma_3}$ are all degenerate: $b_2$ is correlated with $b_{\mathcal{G}_2}$, while $b_{\mathcal{G}_2}$ and $b_{\Gamma_3}$ are strongly anti-correlated. The power spectrum model includes overall 30 independent contributions (28 for the loop-corrections and 2 for the shot noise), in which the bias parameters come as products with various powers and various combinations: for instance, $b_2$ appears as $b_1 b_2$, $b_2^2$, $b_2 f$, so that it can be constrained with power spectrum data only, while $b_{\Gamma_3}$ appears only twice and only linearly in combination with $b_1$ and $f$, which makes it harder to constrain. However, the inclusion of the tree-level bispectrum solves this issue, since the bispectrum model does not include $b_{\Gamma_3}$. Additionally, due to the absence of $b_{\Gamma_3}$, $b_2$ and $b_{\mathcal{G}_2}$ are better constrained since the strong degeneracy is not present anymore. Therefore, we expect measurements of $b_2$ and $b_{\mathcal{G}_2}$ to be more precise from bispectrum measurements only, although $b_1$ is better constrained by power spectrum only measurements.
    \item The noise parameters are anti-correlated with one another since they both control the deviation from Poisson shot noise (constant and scale-dependent). They dominate on small scales together with other effects like FoG and the monopole counterterm, hence we expect correlations between $\alpha_P$,  $\epsilon_{k^2}$ and $\tilde{c}_0$, $c_{\nabla^4 \delta}$.
\end{enumerate}

\section{MCMC analysis}
\label{sec:MCMC}

\subsection{Data}
\label{subsec:data}

In this work we analyse halo power spectrum and bispectrum measurements in redshift space from the large set of 298 Minerva N-body simulations \citep{FirstMinerva, Lippich2019}. These follow the evolution of $1000^3$ dark matter particles in a cubic box of side $L = 1500$ Mpc $h^{-1}$, corresponding to a total effective volume of $\sim 1,000$  Gpc$^3$ $h^{-3}$. The mass cut in the halo catalogues is set to $1.12 \cdot 10^{13} \, M_\odot \, h^{-1}$, with a resulting halo mean number density of $\bar{n} = 2.13 \cdot 10^{-4}$ $h^3$ Mpc$^{-3}$. We limit this study to a single redshift $z=1$. The binning scheme we apply is given by: $k_i = (2+(i-1)) k_f$, with the fundamental frequency $k_f=2\pi L^{-1}$, the bin size is $\Delta k = k_f$, and the total number of bins is $N_k =128$ for the power spectrum and $N_k = 29$ for the bispectrum (with a total number of triangles $N_t = 2766$). For a detailed description of the measurements, the halo catalog construction and the estimators, we refer the reader to \citet{MinervaPSBReal, MinervaBispectrumReal, MinervaBispectrumRSD}.

The fiducial cosmology of the simulations is the $\Lambda$CDM best-fit of the combined analysis of WMAP and BOSS DR9 results \citep{MinervaCosmology} and is given by: $h = 0.695$, $\Omega_\mathrm{m} = 0.285$, $\Omega_\mathrm{b} = 0.046$, $n_s = 0.9632$, $\sigma_8 = 0.828$. This means that the fiducial IDE parameters we should recover are $w=-1$ and $A = 0$.

\begin{figure*}
	\centering
	\includegraphics[width=0.35\textwidth]{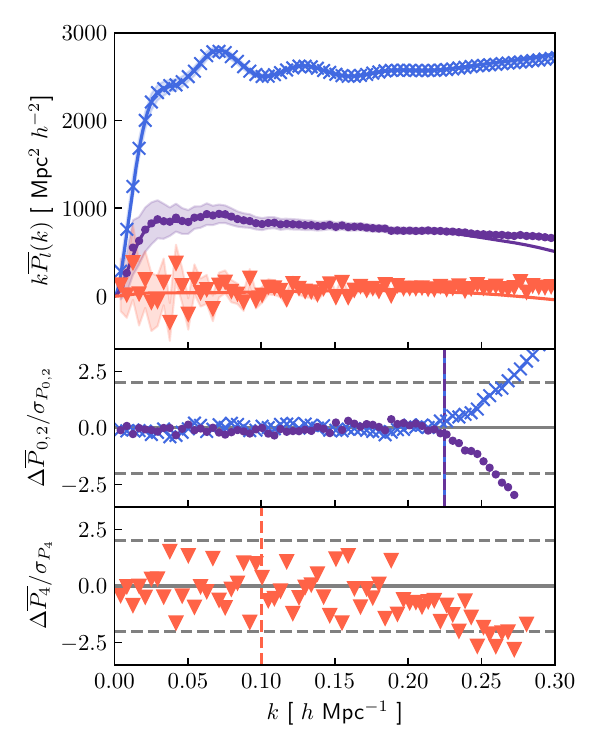}
	\includegraphics[width=0.62\textwidth]{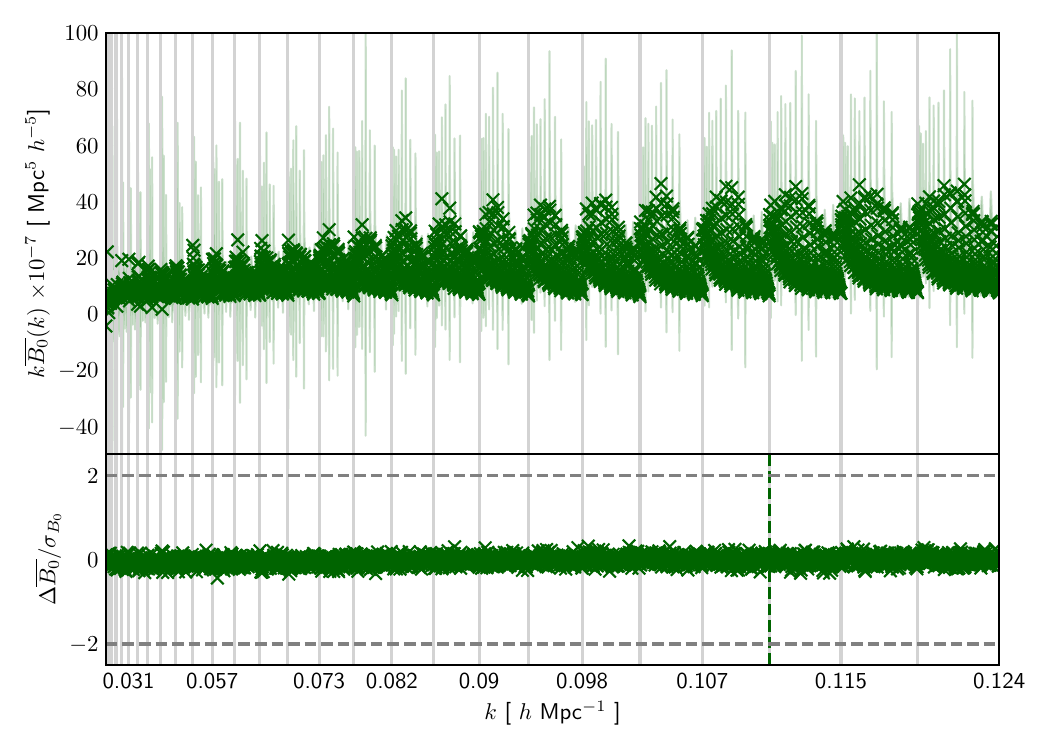}
    \caption{\textit{Left panel}: measurements of the halo power spectrum multipoles -- monopole ($l=0$, blue crosses), quadrupole ($l=2$, purple dots) and hexadecapole ($l=4$, orange triangles). The solid curves correspond to the posterior-averaged theoretical predictions of the joint model with IDE in the following scale-ranges: $k^{l=0,2}_{\mathrm{max}, P} = 0.225$ $h$ Mpc$^{-1}$, $k^{l=4}_{\mathrm{max}, P} = 0.1$ $h$ Mpc$^{-1}$ and $k^{l=0}_{\mathrm{max}, B} = 0.111$ $h$ Mpc$^{-1}$. \textit{Right panel}: measurements of the halo bispectrum monopole (green crosses). In both panels, the average over the 298 N-body simulations is shown. Note that the Poisson shot noise contribution is not subtracted. Shaded areas denote the uncertainties on the mean of the observables. The lower panels show the ratio of residuals to errors on the mean of the observables. The large scattering in the residuals of the hexadecapole is a consequence of the effective binning scheme we adopt in the theoretical prediction. The dashed vertical lines correspond to the values of $k_\mathrm{max}$ used in the fit.}
    \label{fig:data}
\end{figure*}

\subsection{Covariance}
\label{subsec:cov}

The covariance matrix is constructed from 10,000 mock halo catalogues produced with the Pinocchio code \citep{Pinocchio, Munari2017}, 298 of which have initial conditions chosen to match the initial conditions of the 298 Minerva simulations, while the rest have different initial seeds. The total halo power spectrum of the mock catalogs is matched to the one of the N-body simulations at large scales by adjusting the mass threshold \citep{MinervaBispectrumReal}. The large number of mocks allows us to suppress the noise in the off-diagonal components, giving a more accurate covariance, which allows us to better assess the goodness of the fit of the theoretical models we study.

The uncertainties associated to a measurement inferred from this covariance correspond to the volume of a single realisation. However, in our analysis we fit the 298 realisations simultaneously by summing up their log-likelihoods (see \autoref{subsec:like}), thus we effectively fit for the full simulation volume of $\sim 1000$ Gpc$^3$ $h^{-3}$. This volume is unrealistically large when compared to a typical redshift survey, but can be used to study the theoretical systematics of the model. To perform a more realistic analysis, the volume can be decreased in two ways:
\begin{enumerate}
    \item Fitting fewer measurements: it is sufficient to decrease the number of measurements (e.g., in our case 4 realisations correspond to $13.5$ Gpc$^3$ $h^{-3}$), while using the same covariance for a single measurement. However, during various tests we found that the resulting parameters and the goodness-of-fit vary substantially depending on the particular set of simulations selected. We interpret this as a consequence of the fact that the volume of a single simulation is large, since we obtain large amount of low wave-number modes, which are heavily affected by cosmic variance.
    \item Rescaling the covariance: this mimics the effect of analysing a subset of measurements, but with the same suppression in cosmic variance as if we were fitting all of them. Unfortunately, due to the artificial rescaling of the covariance we cannot use the traditional measure of goodness-of-fit, the $\chi^2$-statistic, because the increased error bars do not match the scatter in the measurements. 
\end{enumerate}

In order to mimic the error budget of Stage-IV spectroscopic galaxy surveys, we rescale the measured covariance $C_{ij}$ by a factor $\eta$:
\begin{equation}
\label{eq:cov_resc}
	C_{ij}^\mathrm{resc} = \frac{\eta}{N_\mathrm{R}+ N_\mathrm{C} } \sum_{\alpha=1}^{N_\mathrm{R}+ N_\mathrm{C}} (X^\alpha_i-\overline{X_i})(Y^\alpha_j-\overline{Y_j}) \, ,
\end{equation}
where $N_\mathrm{R}$ and $N_\mathrm{C}$ are the number of measurements from the simulations and catalogs, $X, Y \in \{ P_{l_1}, B_{l_2} \}$ with $l$ being the multipole order, $\alpha$ is the realisation index, $i$ and $j$ denote indices of the wave-number or triangle configuration bin. This factor $\eta$ is a ratio between the effective volumes of all simulation realisations and a typical Stage-IV survey with $V \sim 8$ Gpc$^3 \, h^{-3}$ at $z=1$, giving  $\eta\approx 126$. The effective volume is computed as \citep{EffectiveVolume}
\begin{equation}
    V_\mathrm{eff} = \left[ \frac{\bar{n} P_{hh}(k^*)}{1 + \bar{n} P_{hh}(k^*)} \right]^2 V \, ,
\end{equation}
where $\bar{n}$ is the mean number density, $V$ is the volume of the measured simulation and $k^* = 0.1 \, h$ Mpc$^{-1}$ is the reference scale.
This allows us to roughly match our simulation data to that expected from future surveys, giving similar constraining power and signal-to-noise ratio. It also allows us to extend the range of validity of the theoretical model up to higher values of $k_\mathrm{max}$ with respect to the cuts we would have to adopt in the case of the non-rescaled covariance and the full volume of the simulations. 

It is important to notice that shot noise is not subtracted from our measurements (both the Minerva and the Pinocchio ones). This is particularly prominent in the power spectrum monopole, as can be seen at small scales in  \autoref{fig:data}. In this figure we show the average of the measurements with the corresponding error bars from the rescaled covariance, the chain averaged model from the joint Bayesian analysis (described below) and their residuals. Notice that the very small scatter in the residuals of the monopole and quadrupole (middle-left panel) is an artifact of our inflated errorbars, while for the hexadecapole (bottom-left panel) we have some residual scatter due to the fact that we evaluate the model on the effective modes rather than performing the full average binning.

\subsection{Likelihood evaluation}
\label{subsec:like}
In a Bayesian framework \citep{Bayes:63, MacKay2003}, the probability of a model given some data is proportional to the product between the probability of obtaining the data under the assumption that the model is correct and the probability of the model parameters to have the corresponding values. In other words, the posterior distribution is proportional to the likelihood function multiplied by the prior distribution. 

In our analysis we choose the priors on nuisance parameters to be non-informative: they are uniform and very broad. We consider as fiducial values for these parameters the best fit values obtained in previous studies \citep{MinervaPSBReal, MinervaBispectrumRSD}. Tests were performed to verify that the prior choice does not influence the parameter inference process. However, the priors on the IDE parameters $w$ and $A$ are informative and depend on each other. The equation of state parameter $w$ has a flat prior with size $[-2, -0.5]$. For $w<-2$ dark energy starts to dominate over matter at a later epoch than our redshift of interest $z=1$, so that the growth rate converges quickly to 1 independent from $\xi$ as in the matter-dominated epoch. As for the upper bound, we have observational evidence of the late time acceleration, which implies $w<-1/3$. The parameter $A$ also has a flat prior with size $[-20, 20]$, but with the additional condition that $A$ must have the same sign as $1+w$. The upper limit for $A$ is given under the consideration of the upper bound for $w$ and the most extreme value of $\xi = 50$ b GeV$^{-1}$ tested out in simulations \citep{IDESimulations2015}, while the lower limit is chosen to have a symmetric prior around the fiducial value of $A=0$. Later, when varying additional cosmological parameters, we expand the prior on $A$ to $[-500, 500]$ b GeV$^{-1}$. The additional condition on the sign is explained by the fact that the parameter $\xi = A/(1+w)$ has to be non-negative, since it represents the ratio of two positive quantities: the cross section of dark sector interactions and the mass of dark energy particles (see discussion at the end of \autoref{sec:DE_model}).

Given $N_\mathrm{R} = 298$ independent realisations, we evaluate the total likelihood as the product of individual likelihoods, under the assumption that each is well described by a multivariate Gaussian. As already mentioned, instead of fitting the mean of all realisations with the total volume covariance, we fit each individual simulation and compute
\begin{align}
\label{eq:Ltot}
    &-2 \ln{\mathcal{L}_\mathrm{tot}} =
    -2 \sum_\alpha^{N_\mathrm{R}} \ln{\mathcal{L}_\alpha} = 
    -\sum_\alpha^{N_\mathrm{R}} \chi^2_\alpha = \nonumber\\
    &\sum_{\alpha=1}^{N_\mathrm{R}} \sum_{i, j =1}^{N_\mathrm{b}} \left( X_i^\alpha - X^\mathrm{theo}_i  \right) C_{X, ij}^{-1}  \left( X_j^\alpha - X^\mathrm{theo}_j  \right) \, ,
\end{align}
where $X^\alpha_{i}$ is the power spectrum or bispectrum multipole measurement from the $\alpha$-realisation of the $i$-th Fourier bin from the total number of bins $N_\mathrm{b}$, $X^\mathrm{theo}_i$ is the theoretical prediction and $C_{ij}$ is the rescaled covariance matrix from equation \autoref{eq:cov_resc}. 
We also test applying the correction to the likelihood function suggested in \citet{LikeCorrect}, which should take into account the finite number of mocks used in the construction of the covariance matrix:
\begin{equation}
    -2  \ln{\mathcal{L}_\alpha} = N_\mathrm{M} \ln{ \left(1+\frac{\chi^2_\alpha}{N_\mathrm{M}-1} \right)} \, ,
\end{equation}
where $N_\mathrm{M}$ is the number of mock catalogs.
We found no impact on the posterior distributions, which confirms our expectation of a negligible effect given the large number of mocks catalogues used.

Our likelihood pipeline uses the \texttt{emcee} package \citep{emcee} to sample the posterior distribution. We run the MCMC with 300 walkers, and assume them to have reached convergence after 30000 steps in the analysis which only includes the power spectrum multipoles, and after 15000 steps when the bispectrum multipole(s) is (are) included.

\subsection{Model evaluation}

The evaluation of the theoretical model should in principle be performed on the same grid that is used for the measurements, and the model should be binned in the same way. This can be understood as the resolution of the following discrepancy: the theoretical prediction assumes an infinite universe with an infinite number of $k$-values, while in our case the measurements are performed in boxes with periodic boundary conditions, hence over a restricted number of $k$-bins. In general, there are four approaches to resolve this:
\begin{enumerate}
    \item Full binning: compute an exact average of the theoretical predictions for the observables over each bin in Fourier space. In other words, calculate the theoretical model at each $q_i \in k_i$, with $k_i$ being a Fourier bin and $q_i$ denoting all discrete wave-numbers in a bin of size $\Delta k_i$, and then find a mean value. In our work the bin size equals the fundamental frequency $\Delta k = 2\pi / L$.
    Advantage: this is the most consistent way; disadvantage: it is computationally demanding especially when re-computing the model at each step of the likelihood evaluation.
    \item Effective binning: evaluate the theoretical predictions only at one ``effective'' wave-number, which is computed as the average over each bin centered at $k$ for the power spectrum multipoles, while for the bispectrum the ``effective'' triplet is computed in a similar but hierarchical way \citep[see equation B.2 in][]{MinervaBispectrumRSD}. Advantage: less computationally expensive; disadvantage: it might introduce systematic errors comparable to the statistical uncertainties for the full volume of our data set \citep{MinervaBispectrumReal}.
    \item Expansion approach \citep{MinervaPSBReal}: this is an extension of the effective approach which includes a Taylor expansion of the theoretical model around the effective wave-number up to second order. Advantage: high accuracy, especially for the hexadecapole of the power spectrum; disadvantage: it can be computationally demanding if a large number of terms is involved. 
    \item Integration approach: similar to the averaging approach, but the integral is computed instead of building an average as a sum over the multipoles, divided by the number of all discrete vectors in a bin. This approach is popular in the literature \citep[e.g.,][]{EFTIvanovBispectrum, IntegrationBinning} but similarly to the effective binning it can introduce non-negligible systematic errors, especially for folded triangles. However, this issue can be bypassed by introducing ``discreetness weights'' as corrections \citep{EFTIvanovBispectrum}.
\end{enumerate}
In our study we choose the effective binning for power spectrum and bispectrum multipoles. Our tests show that this type of binning does not have an impact on the inferred parameters due to the enhanced size of our uncertainties after rescaling the covariance, while being the most computationally efficient. 

We also note that in the bispectrum case we include all triangular configurations with $k_1 \geq k_2 \geq k_3$ in our analysis, and do not separate them into squeezed/equilateral/isosceles configurations as is common in the literature. Previous studies \citep[e.g.,][]{MinervaBispectrumReal, IntegrationBinning, SNratio} show that the signal-to-noise ratio and the accuracy of a model increase for high $k_\mathrm{max}$ if nearly equilateral triangle bins are excluded, while still including elongated triangle configurations. Whether or not this selection improves our constraints of dark energy models when adopting realistic error bars is a subject of our future studies. However, we expect the effect to be minimal. 

\begin{figure*}
	\centering
	\includegraphics[width=0.8\textwidth]{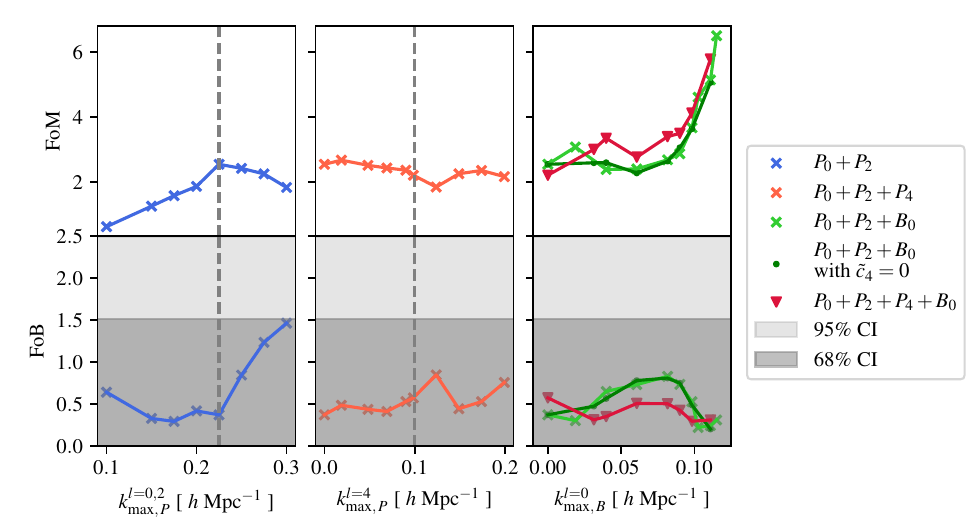}
    \caption{Performance measures as a function of the maximum wave-number of the observable(s) for the following models. \textit{Left panel}: the model includes only the monopole and quadrupole of the power spectrum (blue crosses). \textit{Middle panel}: the hexadecapole (orange crosses) is added to the lower order power spectrum multipoles with their highest Fourier mode fixed to the best scale cut $k^{l=0,2}_{\mathrm{max}, P} = 0.225$ $h$ Mpc$^{-1}$ from the left panel. \textit{Right panel}: the joint models --- the power spectrum monopole and quadrupole with the bispectrum monopole (light-green crosses), the power spectrum monopole and quadrupole plus the bispectrum monopole and one counterterm parameter set to zero (dark-green dots), all power spectrum multipoles and the bispectrum monopole with the highest Fourier mode for the hexadecapole $k^{l=4}_{\mathrm{max}, P} = 0.1$ $h$ Mpc$^{-1}$ (dark-red triangles). The dashed grey lines correspond to the values of $k_\mathrm{max}$ used in the joint analyses: for the power spectrum monopole and quadrupole (left panel) and the hexadecapole (middle panel). The confidence intervals for FoB are calculated by direct integration of a two-dimensional Gaussian over an ellipse between $\pm$ FoB and equating the integral to the 68\% and 95\% percentile thresholds, which results in FoB equal to 1.52 and 2.49, respectively.
    }
    \label{fig:kcuts_main}
\end{figure*} 

\subsection{Model selection}

Our goal is to determine the appropriate scale cuts and forecast the constraining power of future surveys for the IDE parameters, $w$ and $A$. It is crucial to determine the largest range of scales which can be used to extract information from the data in an unbiased manner and tighten the cosmological constraints. This can be also achieved by removing or alleviating degeneracies between the parameters of the theoretical model. For instance, bias relations (i.e., the relations between bias parameters based on their connection due to the underlying physics of gravitational collapse) are able to reduce the parameter space without introducing bias in the estimated parameters. In order to evaluate whether a certain scale cut or a reduction of degeneracies in parameter space provide a better description of the observables we need a way to quantitatively compare the posterior distributions from our Markov chains. Since the $\chi^2$-statistic is not valid anymore as a measurement of the goodness-of-fit, due to the rescaling of the covariance matrix (see discussion in \autoref{subsec:cov}), we make use of two performance metrics to select the best scale cut and/or reduced model \citep[see a similar approach in][]{TestinBiasNkNoise_bG2rel, PerformMetrics}:

\begin{enumerate}
	\item the Figure of Bias (FoB) quantifies the relative separation of the measured parameters from their fiducial values in terms of the variance of the posterior distribution. 
	We calculate the 68\% - 95\% percentile thresholds for the FoB by assuming the posterior distribution of the parameters of interest to be Gaussian. Hence, for one parameter the 68\% threshold equals 1, while for two parameters the integration of two-dimensional Gaussian over an ellipse results in a FoB of 1.52.  It is defined as 
	\begin{equation}
		\mathrm{FoB} = \sqrt{(\theta_\mathrm{fid} - \bar{\theta}) S^{-1} (\theta_\mathrm{fid} - \bar{\theta})} \, ,
	\end{equation}
	where $\bar{\theta}$ and $\theta_\mathrm{fid}$ are vectors labeling, respectively, the posterior averages and fiducial values of the parameters we want to measure, and $S = \mathrm{cov} (\theta)$ is the covariance matrix of the parameters calculated from the chains. In our case, $\theta = [w, A]$  and $\theta_\mathrm{fid} = [-1, 0]$;
	
	\item the Figure of Merit (FoM) shows the merit of the considered model with respect to the parameters varied in the fit. It is the inverse of the volume of the 68\% contours of the parameters, effectively giving a global measure of the inverse size of the uncertainties obtained on the parameters \citep{Wang:2008zh}. This is given by
	\begin{equation}
		\mathrm{FoM} = \frac{1}{\sqrt{\mathrm{det}(S)}} \, .
	\end{equation}	
\end{enumerate}

\section{Results}
\label{sec:results}

\subsection{Joint base model}

In total our base model contains 13 parameters: \{$w$, $A$, $b_1$, $b_2$, $b_{\mathcal{G}_2}$, $b_{\Gamma_3}$, $\tilde{c}_0$, $\tilde{c}_2$, $\tilde{c}_4$, $\alpha_P$, $\epsilon_{k^2}$, $c_{\nabla^4 \delta}$, $\alpha_B$ \}. Firstly, we aim to find the highest $k_\mathrm{max}$ at which the model parameters are not biased while the constraints on the dark energy parameters are the most informative (i.e., FoM reaches its maximum). For this purpose we run our likelihood pipeline, calculate the performance metrics and check the posterior distributions for the nuisance parameters. The latter is done in order to avoid any deviation larger than two standard deviations from the fiducial values for $b_1$, $b_2$, $b_{\mathcal{G}_2}$, $b_{\Gamma_3}$ $\alpha_P$, $\epsilon_{k^2}$ and $\alpha_B$ as determined in the previous analyses of \citet{MinervaBispectrumReal, MinervaPSBReal, MinervaBispectrumRSD}. Note that the counterterms in real space and in redshift space are not the same.

We vary the value of $k_\mathrm{max}$ jointly for the power spectrum monopole and quadrupole, as is typically done in analyses of observational data \citep{FlorianMultipoles}. The scale cut is varied in the range of $k^{l=0, 2}_{\mathrm{max}, \, P} \in [0.1, 0.3]$ $h$ Mpc$^{-1}$ and is found to maximize the FoM while keeping the FoB within one standard deviation at $0.225$ $h$ Mpc$^{-1}$. This can be seen in the left panel of \autoref{fig:kcuts_main}, where blue crosses correspond to the joint analysis of the lower order power spectrum multipoles. Keeping this value fixed, we then add the hexadecapole with $k^{l=4}_{\mathrm{max}, \, P} \in [0, 0.2]$ $h$ Mpc$^{-1}$. We expect our model to describe it less efficiently than the lower order multipoles, therefore the range of scales included for $P_4$ is smaller than for $P_0$ and $P_2$. Unsurprisingly the addition of the hexadecapole allows us to constrain $\tilde{c}_4$ and hence improve significantly the constraints on $\tilde{c}_2$, as well as on the other counterterms, though less significantly. However, we see in the middle panel of \autoref{fig:kcuts_main} no improvement in the constraints of the IDE parameters, the hexadecapole only slightly decreases the FoM and increases the FoB. This finding is in agreement with the conclusions of \citet{PedroCola}.

Next we include the bispectrum monopole, $B_0$. This is shown in the right panel of \autoref{fig:kcuts_main}. We perform the analysis by adding it to $P_0+P_2$, since the power spectrum hexadecapole does not contribute to the improvement of the  constraints in the IDE parameters. The results are denoted by the light green crosses connected by a solid line. We observe an increase in the FoB and no improvement in the FoM for scales up to $k^{l=0}_{\mathrm{max}, \, B} = 0.08$ $h$ Mpc$^{-1}$ for $P_0+P_2+B_0$. However, for $k^{l=0}_{\mathrm{max}, \, B} > 0.08$ $h$ Mpc$^{-1}$ we notice a steep growth in the FoM as we add more $k$-bins. Clearly, inclusion of the small-scale information contained in the higher $k$-bins provides better constraining power, especially since the number of triangles increases significantly with each bin. The results do not show a bias in any of the parameters of the model, at least up to the 27-th $k$-bin (corresponding to $k^{l=0}_{\mathrm{max}, \, B} = 1.12$ $h$ Mpc$^{-1}$) from the total 29 bins of our measurements. 

Additionally, we perform a joint analysis with the hexadecapole of the power spectrum included up to $0.1$ $h$ Mpc$^{-1}$, which is rather conservative and common in the literature. This model with all power spectrum multipoles and $B_0$ shows a mild but continuous improvement of the FoM in comparison to the model with only lower order power spectrum multipoles (see the dark red triangles in the right panel). Noticeably, adding the power spectrum hexadecapole in the joint analysis with the bispectrum monopole always results in a smaller FoB (and thus bias) of the IDE parameters. We check whether the same effect could be achieved by setting the otherwise unconstrained $\tilde{c}_4$ to zero in the model with the monopole and quadrupole only. As expected, the constraints of the $P_0+P_2+B_0$ model are not sensitive to this change. This is shown by the dark-green dots in the right panel, which overlap with the light-green crosses of the base model with $\tilde{c}_4 \neq 0$. We therefore conclude that the inclusion of the hexadecapole of the power spectrum, in combination with the bispectrum monopole, can help (especially at lower scale cuts) by breaking degeneracies between the counterterms and bias parameters. 

\begin{figure}
	\includegraphics[width=\linewidth]{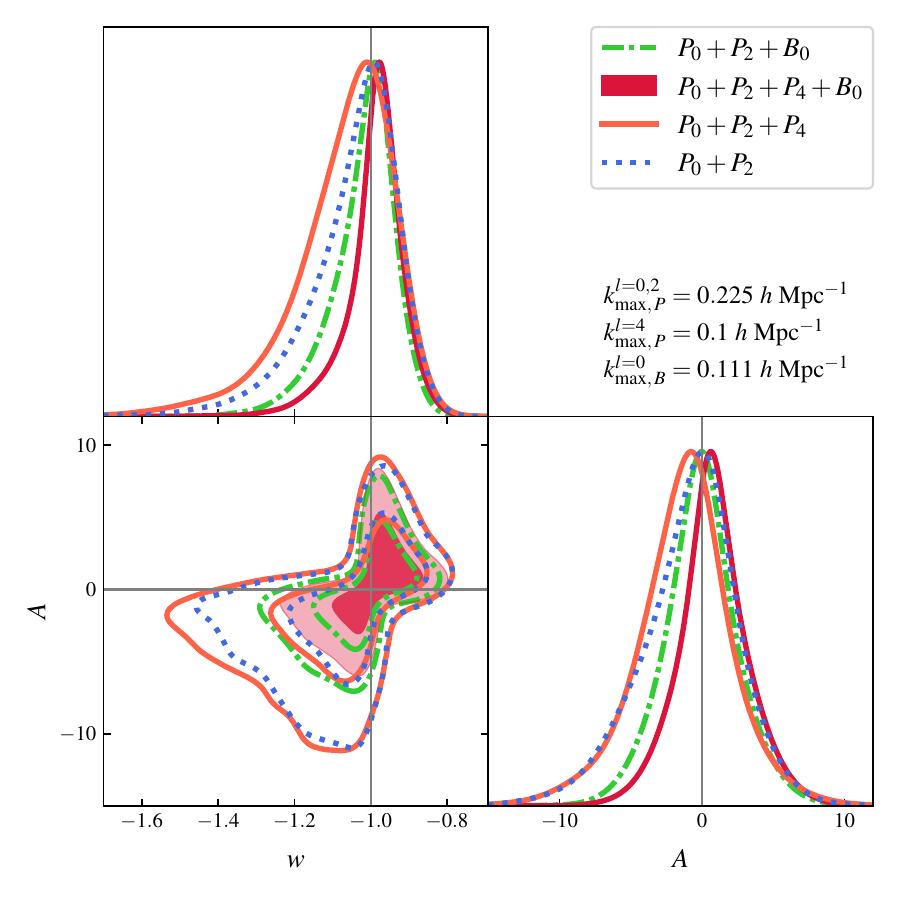}
    \caption{Marginalized posterior distributions for the IDE parameters in the base model and with the scale cuts as specified on the plot. The power spectrum monopole and quadrupole analysis is denoted by the dotted blue line, constraints from all power spectrum multipoles are given by the solid orange line, the joint analysis of the power spectrum monopole plus quadrupole and bispectrum monopole is presented by the dotted-dashed light-green line, and the full joint analysis is shown with the dark-red line. Clearly, inclusion of the bispectrum monopole improves constraining power by $\sim 30\%$. The thin grey lines correspond to the fiducial values from the $\Lambda$CDM cosmology. Note that parameter $A$ is given in the units of b GeV$^{-1}$. The fact that the contour shows non-vanishing values for the forbidden regions with $A/(1+w)<0$ is an artifact of the smoothing used for plotting. }
    \label{fig:butterfly_main}
\end{figure}

\begin{figure*}
	\centering
	\includegraphics[width=0.8\textwidth]{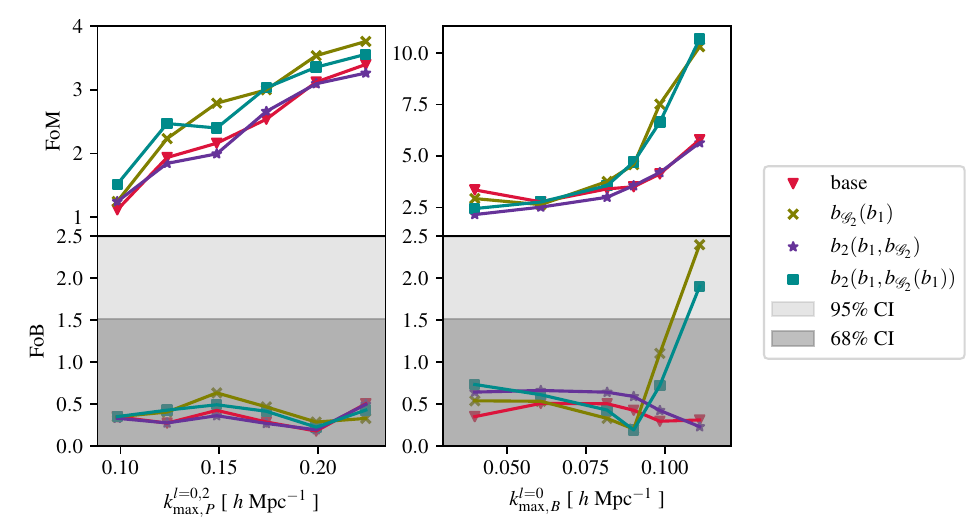}
    \caption{Implementation of the bias relations into the joint analysis of all power spectrum multipoles with the bispectrum monopole. \textit{Left panel}: performance metrics as a function of the maximum wave-number of the power spectrum monopole and quadrupole with the highest Fourier mode of the bispectrum monopole set to the moderate value of $k^{l=0}_{\mathrm{max}, B} = 0.08$ $h$ Mpc$^{-1}$.  \textit{Right panel}: performance metrics as a function of the maximum wave-number of the bispectrum monopole with the highest Fourier modes of the power spectrum multipoles set to $k^{l=0,2}_{\mathrm{max}, P} = 0.225$ $h$ Mpc$^{-1}$ and $k^{l=4}_{\mathrm{max}, P} = 0.1$ $h$ Mpc$^{-1}$. The dark-red triangles correspond to the base model with 13 parameters, the olive crosses denote the tidal bias relation, the purple stars represent the $b_2$-relation, the dark-cyan squares denote the combined relation of $b_2$ and $b_{\mathcal{G}_2}$. Note that the application of the combined or the tidal-bias relation leads to stronger constraining powers on the IDE parameters. This is especially prominent in the right panel for $k^{l=0}_{\mathrm{max}, B} \geq 0.09$ $h$ Mpc$^{-1}$. The confidence intervals for FoB are calculated as in \autoref{fig:kcuts_main}.}
    \label{fig:kcuts_biasrel}
\end{figure*}

The full posterior plot is shown in  \autoref{fig:base_full} of \autoref{sec:app_full}. Additionally, we present in \autoref{fig:butterfly_main} the triangle plot for the marginalized posterior distribution of the IDE parameters. The characteristic ``butterfly'' pattern, in which the posterior distribution is present only in two quadrants -- upper-right and lower-left, is the consequence of the priors on the IDE parameters  with the condition $A/(1+w) \geq 0$, as described in \autoref{subsec:like}. The $w-A$ contour shows a strong degeneracy: as argued in \autoref{sec:DE_model}, a very negative value of $w$ leads to a later start of the dark energy dominated epoch, hence to a longer matter domination which is associated with increased growth of structures. A negative value of $A$ leads to extra drag in \autoref{eq:Euler}, and thus to an enhancement of the growth of structures in the linear regime (the opposite is true for the nonlinear regime, as discussed in \autoref{sec:DE_model}). On large scales the drag term acts in the same direction as the gravitational acceleration, since the velocity field is always aligned with the spatial gradient of the gravitational potential in the linear regime. The opposite happens for $w>-1$ and $A>0$. 
This is in line with what can already be seen in \autoref{fig:IDEfD} and its diagonal (from upper left to lower right) pattern in the growth rate $f$, and is particularly prominent in the upper right corner (for positive values of $A$ and $(1+w)$). There we observe the same level of suppression for either a larger $w$ value with $A$ fixed, or for a larger $A$ value with $w$ fixed. It is therefore apparent how the two effects are strongly degenerate, which results in larger uncertainties on them, especially in biased cases. Such a conclusion is in agreement with the findings of \citet{PedroCola}.

In addition to the base model with 13 parameters, we run the joint analysis adopting some assumptions that are common in the literature: $b_{\Gamma_3} = 0$ and / or $\epsilon_{k^2} = 0$. Assuming that noise is scale-independent (as opposed to leaving the $\epsilon_{k^2}$ parameter free), results in no improvement and no bias in the constraints. This is a consequence of the rescaled error bars, since we know from the real space analysis with the full volume of the simulations by \citet{MinervaPSBReal} that the scale-dependent term in shot noise is relevant at larger Fourier modes due to accounting for additional corrections beyond the assumed one-loop model. The assumption of $b_{\Gamma_3} = 0$ does not yield improved constraints, but it biases the noise parameters $\alpha_P$, $\epsilon_{k^2}$ and $\alpha_B$ by more than $2\sigma$ with respect to the fiducial values. This is in line with the joint analysis performed on the same data set in real space, which measures $b_{\Gamma_3} = 0$ to be inconsistent with zero at more than $2\sigma$ \citep{MinervaPSBReal}. The assumption of both conditions results in bias in the noise parameters and $b_{\mathcal{G}_2}$ at more than $2\sigma$ level, while $\alpha_B$ gets biased at more than $3\sigma$ level. From this we conclude that, when including the bispectrum monopole, $b_{\Gamma_3}$ should not be set to zero. 

\subsection{Bias relations}

Although the strong degeneracy between $b_{\mathcal{G}_2}$ and $b_{\Gamma_3}$ is broken by the inclusion of the bispectrum monopole in the analysis, we are still interested in the impact of bias relations  for two reasons: firstly, to reduce the dimensionality of the parameter space; secondly, we know that both $A$ and $w$ are correlated with $b_1$, hence a tighter constraint on the latter will lead to tighter constraints on the IDE parameters.  

\begin{figure}
    \centering
	\includegraphics[width=0.8\linewidth]{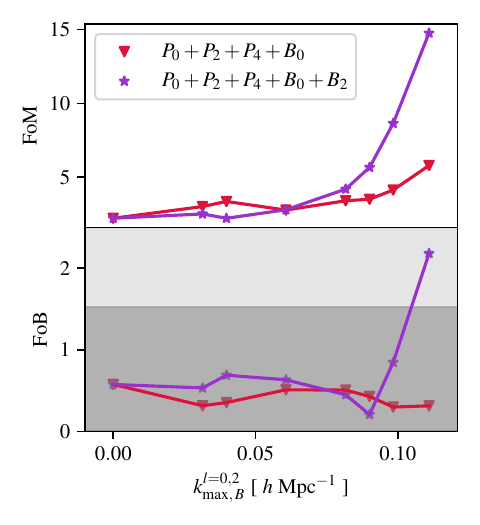}
    \caption{Joint analysis of all power spectrum multipoles with the bispectrum monopole (dark-red triangles) and quadrupole (purple stars). The performance metrics, FoM and FoB, are presented as a function of the maximum wave-number on both multipoles of the bispectrum. Evidently, inclusion of the bispectrum quadrupole leads to tighter constraints on the IDE parameters for $k^{l=0,2}_{\mathrm{max}, B} \geq 0.09$ $h$ Mpc$^{-1}$. The confidence intervals for the FoB are calculated as in \autoref{fig:kcuts_main}.}
    \label{fig:kcuts_B0B2}
\end{figure}

We investigate the full joint analysis of $P_0+P_2+P_4+B_0$ for the following bias relations: 
\begin{enumerate}
    \item the fit from the separate universe simulations presented in \citet{b2rel}\footnote{The additional term $\frac{4}{3} b_{\mathcal{G}_2}$ is due to the difference in the definition of the tidal-bias operator in the reference and our notation.} : 
    \begin{equation}
            b_2(b_1, b_{\mathcal{G}_2}) = 0.412 - 2.143 b_1 + 0.929 b_1^2 +0.008 b_1^3 + \frac{4}{3} b_{\mathcal{G}_2} \, ;
    \end{equation}
    \item the fit from the excursion set approach of \citet{TestinBiasNkNoise_bG2rel}:
    \begin{equation}
    b_{\mathcal{G}_2}(b_1) = 0.524 - 0.547 b_1 + 0.046 b_1^2 \, ;
    \end{equation}
    \item the combination of both relations: $b_2(b_1, b_{\mathcal{G}_2}(b_1))$.
\end{enumerate}
These relations have been shown to improve the fits of the power spectrum and bispectrum in \citet{MinervaBispectrumReal, MinervaPSBReal, MinervaBispectrumRSD}. Results for the individual relations as well as for the combined one are shown in \autoref{fig:kcuts_biasrel} in terms of the FoM and FoB metrics. We notice that the $b_2(b_1, b_{\mathcal{G}_2})$ relation, denoted by purple stars, does not impact our constraints both when varying the maximum wave-number for the power spectrum (left panel) and for the bispectrum (right panel). This can be explained by a combination of two factors. Firstly, the fact that in our base model the $b_2$-relation always stays consistent with the fiducial value of $b_2$, \citep[that measured from the real space analysis of][]{MinervaBispectrumReal, MinervaPSBReal, MinervaBispectrumRSD}, which implies that we are not introducing any new information with the bias relation. Then, while a different behaviour has been seen in the analysis of \citet{MinervaPSBReal} in Fig.~8 with very small error bars, the fact that ours are ``realistic'' (larger) for Stage-IV surveys means that we cannot detect this effect. On the other hand, the tidal bias relation (olive crosses) and its combination with the $b_2$-relation (dark-cyan squares) lead to a reduction in the constraints on $w$, $A$, $b_1$ and $b_2$. This improvement is particularly prominent in the right panel, where the number of modes in the bispectrum monopole is varied while the scale-range of power spectrum multipoles is fixed. At $k^{l=0}_{\mathrm{max}, B}=0.1$ $h$ Mpc$^{-1}$ we see the $b_{\mathcal{G}_2(b_1)}$ has a higher value of FoM, which equals $\simeq 7.5$ and is larger than FoM$\simeq 4$ of the base model with the same largest Fourier mode. However, for $k^{l=0}_{\mathrm{max}, B}>0.1$ $h$ Mpc$^{-1}$ cases (ii) and (iii) cause a continuous shift of the IDE parameters to the positive values of $A$ and $(1+w)$. Nonetheless, even for  $k^{l=0}_{\mathrm{max}, B}=0.111$ $h$ Mpc$^{-1}$ all nuisance and IDE parameters stay within 2 standard deviations of their fiducial values (see the full posterior distribution in \autoref{fig:biasrel_full}), FoM reaches $\simeq 11$ for both relations. From the lower right panel we see that the combined bias relation results in the less biased dark energy parameters, in contrast to the tidal-relation only.  

We note that the bias relations applied here are derived in the context of the $\Lambda$CDM cosmology. From the simulations performed in \citet{IDESimulations2015}, we know that at the redshift of our interest deviations of the halo mass function from the standard cosmology are negligible. Moreover, the measurements in our analysis do not include modifications in the gravitational interaction. Therefore, the application of the bias relations is justified. Nevertheless, we aim to study the impact of the interactions between dark matter and dark matter on the bias relations in our future work.

\subsection{Bispectrum quadrupole}

Recent progress in measuring the bispectrum multipoles from simulations \citep{MinervaBispectrumRSD} showed that, while the monopole constrains mostly the bias parameters $b_1$, $b_2$ and $b_{\mathcal{G}_2}$, the quadrupole can break the $b_1-f$ degeneracy and can constrain the growth factor even in the case of a bispectrum only analysis. These constraints on the growth rate from the bispectrum multipoles are weaker than the ones that can be obtained from the power spectrum multipoles. The argument for the inclusion of higher order multipoles of the bispectrum is also discussed in \citet{BispAniso1, BispAniso2}.

However, it raises the question: if including the bispectrum quadrupole improves the constraints on the growth rate, does it also do so for the IDE parameters, $w$ and $A$? In particular, it is interesting to investigate if this holds also in the case of Stage-IV-like error bars, such as the ones we employ here.
We run our likelihood pipeline for the base model and find that indeed, for $k^{l=0,2}_{\mathrm{max}, B}>0.09$ $h$ Mpc$^{-1}$, the addition of the quadrupole improves the constraints on the dark energy parameters. Our results are shown in \autoref{fig:kcuts_B0B2}, where dark-red triangles denote the power spectrum multipoles plus the bispectrum monopole, while the addition of $B_2$ is denoted by purple stars. We note that the FoB and FoM as functions of $k_\mathrm{max}$ resemble the trend obtained in the previous section with the bias relations, with the exception that in the present case the slope of the FoM in the upper panel is steeper at large wave-numbers. For instance, for $k^{l=0,2}_{\mathrm{max}, B}=0.1$ $h$ Mpc$^{-1}$, we get FoM $\simeq 9$, while with the tidal bias relation from the previous section and without the bispectrum quadrupole we obtained FoM $\simeq 7.5$. The full posterior distribution for the $k^{l=0,2}_{\mathrm{max}, B}=0.111$ $h$ Mpc$^{-1}$ can be seen in  \autoref{fig:B0B2_full}. The increase in the FoM is the result of an improvement in the constraints on the IDE parameters and the linear bias. Moreover, because of the degeneracy between the linear bias and the other bias parameters, a tighter constraint on the former leads to the latter being more constrained too. This effect is more significant than in the case of the bias relations, and avoids resorting to such relations, which might not be valid in extended cosmological models. 
This is a hint that the bispectrum quadrupole might play a crucial role in constraining extended cosmological models with spectroscopic galaxy clustering data.

\begin{figure*}
    \centering
	\includegraphics[width=0.49\textwidth]{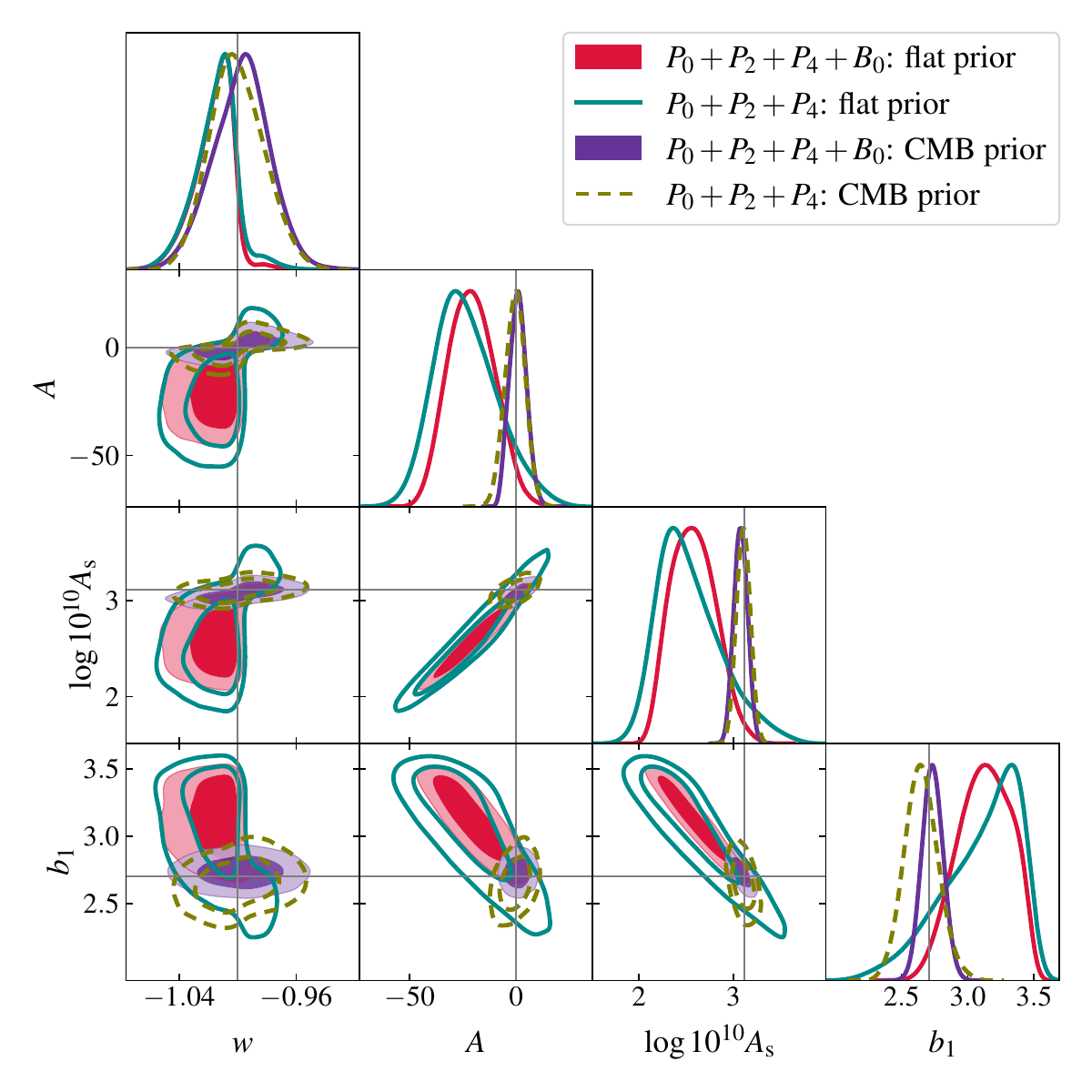}
 \includegraphics[width=0.49\textwidth]{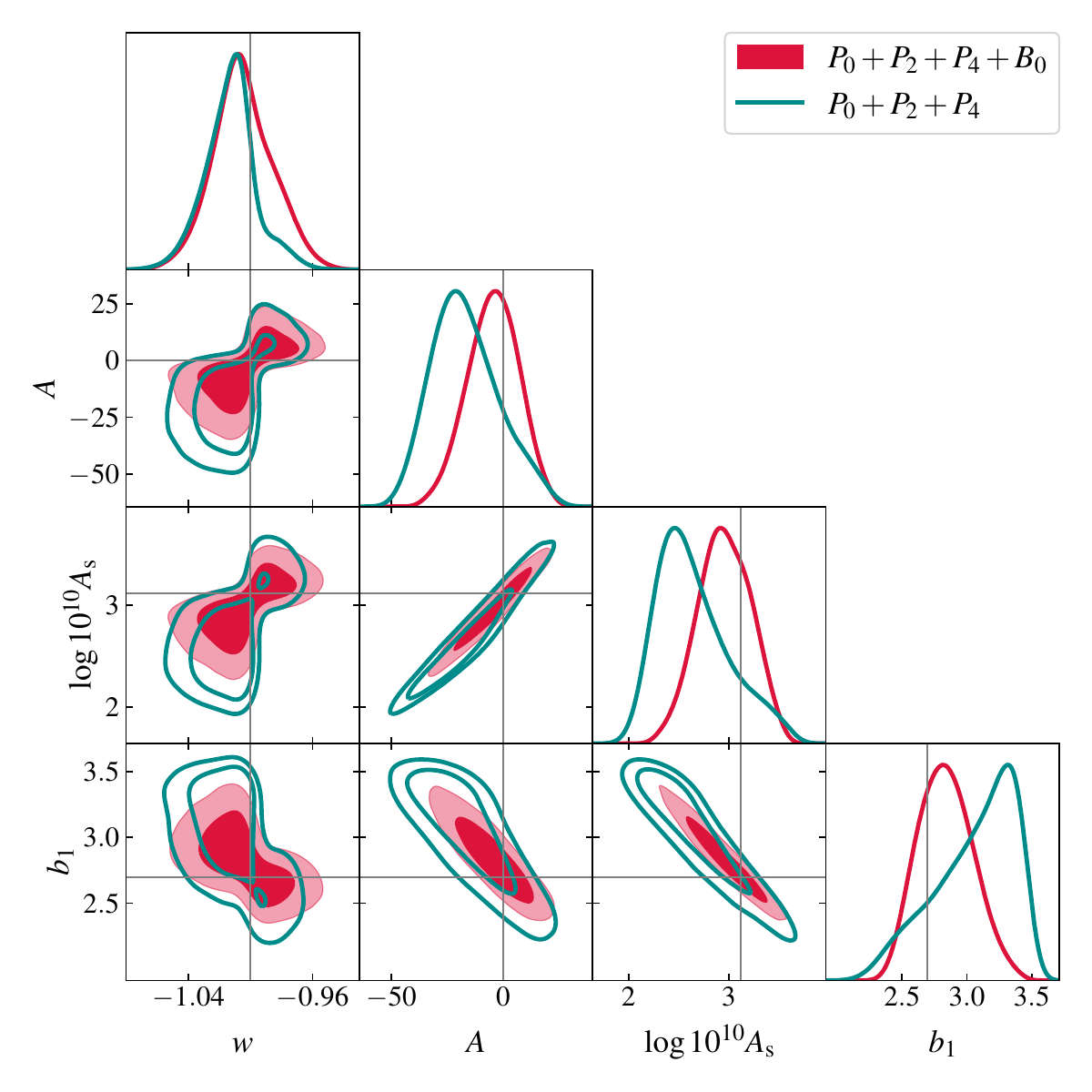}
    \caption{\textit{Left panel}: analysis with the simulated data. Dashed olive lines and purple contours show results with the Gaussian prior on $A_\mathrm{s}$ around its fiducial value with 3 standard deviations from the CMB analysis. \textit{Right panel}: analysis with the synthetic data and flat priors. Grey lines denote the fiducial values. In both cases the AP-effect is mimicked by a tight Gaussian prior on $w$ around its fiducial value and with $\sigma_w = 0.02$. Red contours represent the joint analysis, dark cyan lines represent the analysis with power spectrum multipoles only. Scale cuts are the same as in \autoref{fig:butterfly_main}.}
    \label{fig:wACDM_As}
\end{figure*}

\subsection{Forecast for $w$CDM}

In addition to IDE, we repeat our analysis for the simplest extension to $\Lambda$CDM model in which $w = w_0$ is constant, but can assume values different from $-1$. We assume the same flat prior of $w$ as described in \autoref{subsec:like} without imposing any additional theoretical constraints \citep[similar to][]{GuidowCDM}. However, $w<-1$ leads to gradient instabilities and should not be considered as a viable smooth quintessence theory in which the dark energy speed of sound equals the speed of light. See \citet{DAmico:2020} for the analysis with the theoretical prior on $w$ in smooth quintessence, as well as the case of $w<-1$ in clustering quintessence. 

Based on what we have learned with the IDE model we perform the following analyses:
\begin{enumerate}
    \item base model with $k^{l=0}_{\mathrm{max}, B} = 0.111$ $h$ Mpc$^{-1}$;
    \item combined bias relations with $k^{l=0}_{\mathrm{max}, B} = 0.095$ $h$ Mpc$^{-1}$;
    \item bispectrum monopole and quadrupole with $k^{l=0,2}_{\mathrm{max}, B} = 0.09$ $h$ Mpc$^{-1}$.
\end{enumerate}
We obtain the following constraints on the dark energy equation of state parameter for the three runs respectively: $w =-0.99 \pm 0.10$, $-0.98 \pm 0.10$, $-0.99 \pm 0.09$. The full posterior is shown in \autoref{fig:wCDM_full}. All three cases result in similar level of uncertainty on the inferred parameters, except on the shot noise parameter of the bispectrum for which both the bias relations and the inclusion of $B_2$ yield tighter constraints. However, the improvement with respect to the analysis with the power spectrum only is apparent, roughly $38\%$, demonstrating once again that the inclusion of the bispectrum does increase the cosmological information that can be extracted from the data.

With the same setup of $k^{l=0,2}_{\mathrm{max}, P} = 0.225$ $h$ Mpc$^{-1}$, $k^{l=4}_{\mathrm{max}, P} = 0.1$ $h$ Mpc$^{-1}$ and $k^{l=0}_{\mathrm{max}, B} = 0.111$ $h$ Mpc$^{-1}$ we obtain larger uncertainties on $w$ for the $w$CDM cosmology than in the IDE case. This may seem counter-intuitive, since $w$CDM contains one fewer parameter. However, when we compare posterior distributions against each other, the contours in $w$CDM are always tighter or equal to the IDE case and the difference in $\sigma_w$ can be explained by the non-Gaussian form of the posterior distribution and the corresponding projection effects in the IDE case \citep[for more about projection effects of marginalisation see][]{GomezValent2022}.

\subsection{Degeneracies with cosmological parameters}

We discussed above the degeneracies between nuisance and IDE parameters in the case of fixed cosmology. Now we aim to understand the interplay between $w$, $A$ and cosmological parameters. We first consider the scalar amplitude $A_\mathrm{s}$ with a flat prior $1 \leq \ln{10^{10} A_\mathrm{s}} \leq 4$. Since $w$ and $A$ enter the model through the linear growth factor, they are degenerate with any parameter controlling the amplitude of the power spectrum (as argued above for the degeneracy with the linear bias $b_1$). Hence, there is a strong degeneracy between IDE parameters and the primordial amplitude $A_\mathrm{s}$, which limits the measurement of those parameters individually in an analysis where the full cosmological parameter space is explored.

However, in such a realistic cosmological analysis, one should also take into account the Alcock-Paczynski effect \citep[AP,~][]{Alcock1979} arising from the assumption of a particular cosmology when converting redshifts to distances in the measured data. This results in a re-scaling of the $k$-modes and angles, which is sensitive to $w$ as well as to $\Omega_\mathrm{m}$ and $h$ \citep[see equations A.11 in][]{Ivanov:2020}. Including that in the analysis, we find that, even when varying only the nuisance parameters in addition to the dark energy parameters, taking into account the AP-effect tightens the constraint on $w$ by nearly 5 times (see \autoref{fig:AP_full} in Appendix)\footnote{The analysis in this section is performed with the \texttt{pocoMC} sampler \citep[]{karamanis2022pocomc,karamanis2022accelerating} with the standard convergence criterion, $10^3$ live points and $3 \times 10^4$ additional sampling points. We switch to this MCMC sampler from \texttt{emcee} after reproducing results from the previous sections with the new sampler. This change is justified by the fact that this sampler is more computationally efficient and better suited to explore a parameter space with strong degeneracies.
Moreover, we compute the linear power spectrum with the \texttt{bacco} linear emulator \citep[]{bacco2021} instead of calling a Boltzmann solver at each step of the chain, which significantly speeds up the analysis.}. 

We then perform the analysis with the inclusion of the AP-effect and varying $A_\mathrm{s}$ with the power spectrum multipoles only, and find a strong positive correlation between $A_\mathrm{s}$ and $A$ and a negative one with $b_1$ \citep[analogous to the findings of][e.g., Figure 7 and 18]{Carrilho:2022}. This can be understood as follows: a very negative value of $A$ corresponds to more power at linear scales, which requires a smaller value of the magnitude $A_\mathrm{s}$ and larger value of the linear bias $b_1$ to compensate. As a result, we find that the AP-effect confines the equation-of-state parameter to $w = -1.03 \pm 0.02$. We use this as a tight Gaussian prior on $w$ in the following analysis. Note that a significant shift from the fiducial values is present only in $w$, $A$, $A_\mathrm{s}$ and $b_1$, while the rest of the nuisance parameters remain unbiased and are not dominated by priors.

Next we perform an analysis with a tight Gaussian prior on $w$ around its fiducial value in order to mimic the AP-effect and study whether the addition of the bispectrum monopole can break the strong degeneracy between $A$ and $A_\mathrm{s}$. The reason for avoiding a direct implementation of the AP-effect for the bispectrum is the high computational demand when taking into account all possible triangle configurations, a solution for which we leave for future work. As can be seen in the left panel of \autoref{fig:wACDM_As}, the addition of the bispectrum (red contours) to the power spectrum multipoles (dark cyan lines) makes the marginalised posterior distributions more Gaussian, reduces constraints on $A$ and $A_\mathrm{s}$ by $~25-27\%$, while bringing the mean and best-fit values closer to the fiducial values of these parameters. Constraints on the linear bias $b_1$ obtained with the power spectrum multipoles only are very broad, and prefer values close to the upper limit of its prior $[0.9, 3.5]$. Should this be increased, the degeneracy would drive $b_1$ to higher values, $A_\mathrm{s}$ to even lower values, and make $A$ more negative. Analogous to findings from the previous sections, the bispectrum improves constraints on all bias parameters, including $b_1$, and shifts its best-fit value closer to the fiducial one. As a consequence, it drives the peaks of the $A$ and $A_\mathrm{s}$ distributions to their true values too. A similar effect is seen for fixed $A_\mathrm{s}$ in \autoref{fig:butterfly_main}: with the power spectrum multipoles only, there is a clear asymmetry in the 2-D ``butterfly'' towards negative values of $w$ and $A$, while adding the bispectrum monopole visibly improves the symmetry of the contour.

Although we recover the true values within 2 standard deviations, the degeneracy between $A$ and $A_\mathrm{s}$ still leads to a preference of lower values for $A_\mathrm{s}$ and $A$. This degeneracy is decreased if one applies a Gaussian prior on $A_\mathrm{s}$, centered on its fiducial value and with a size of 3$\sigma$ from the WMAP+BOSS DR9 results \citep{MinervaCosmology}. We can impose a CMB-based prior because the IDE model introduces deviations to $\Lambda$CDM only in the late Universe, thus results obtained from CMB data assuming $\Lambda$CDM would still be valid. In this case, $w$ fills in the prior without preferring values smaller than $-1$, $A_\mathrm{s}$ is fully defined by its prior as well, and the $~16\%$ improvement in constraints on $A$ after adding the bispectrum monopole (purple contours) to the power spectrum multipoles (olive dashed lines) is explained by the $~38\%$ shrinking in the constraints on the linear bias $b_1$.

The fact that a strong degeneracy leads to a bias in the parameters hints at the existence of projection effects of the marginalisation. In order to verify this, we create synthetic data for the same fiducial cosmology, using the best-fit values obtained from the $\Lambda$CDM fits for the nuisance parameters. As before, we mimic the AP-effect by setting a tight prior on $w$ around its true value. As can be seen from the results in the right panel of \autoref{fig:wACDM_As}, a flat prior on the primordial scalar amplitude results in biases towards negative and smaller values in $A$ and $A_\mathrm{s}$ respectively, which are similar to the biases appearing in the analysis of the simulated data on the left panel of the same figure. Also analogously to the results with data from the simulations, in the analysis with the power spectrum multipoles only, the linear bias $b_1$ tends to the upper limit of its prior range. However, the addition of the bispectrum monopole allows to recover the fiducial values and reduces the degeneracy and the biases due to marginalisation, showing that the bispectrum is a promising tool for improving this type of projection effects.

It is important to note that, in spite of projection effects explaining some of the biases, the joint analysis on simulation data still favours a more negative value for $A$ and $w$ than the same analysis on synthetic data. We know that these biases are not caused by the invalidity of the model at the chosen scale cuts when the parameter space is opened to the additional cosmological parameters. Tests in $\Lambda$CDM and more importantly $w$CDM scenarios when $A_\mathrm{s}$ is varied do not demonstrate any significant deviations from the fiducial values. The actual reason for this is the following. As can bee seen from the orange lines in \autoref{fig:butterfly_main}, the model already slightly prefers the negative values of $A$ and $w<-1$. This implies that the presence of a strong degeneracy on top of even slight theoretical errors (discrepancies between a theory and data) amplifies the effects from the latter. Similar conclusions can be found in the analysis for DGP gravity in \citet{Piga2022}.    \\

We now discuss the degeneracies appearing when varying additional cosmological parameters. Two other cosmological parameters enter the Euler equation (\autoref{eq:Euler}) and the AP-effect through the Hubble function: the fractional energy density of non-relativistic matter $\Omega_\mathrm{m}$ and the expansion rate $h$. Our tests with the power spectrum multipoles only, including the AP-effect (see \autoref{fig:wACDM_AP_full}) agree with the results from \citet{Carrilho:2022}, showing the degeneracy between these parameters and the IDE parameters is not as substantial as that with the scalar amplitude $A_\mathrm{s}$. Parameters $w$ and $h$ are strongly degenerate, since both of them affect the expansion history: smaller $h$ or very negative $w$ both lead to smaller Hubble parameter; the same holds for $w$ and $\Omega_\mathrm{m}$, though the degeneracy is less severe. We obtain an estimate of the expansion rate $h$ biased towards higher values, which is a consequence of its degeneracy with $w$. For the same set-up in the $\Lambda$CDM scenario this effect is not present. This degeneracy between $w$ and $h$ can be broken by combining data from different redshifts or from BAOs \citep[e.g., as in][]{GuidowCDM}. BAO data also constrain $\Omega_\mathrm{m} h^2$, which breaks its degeneracy with the spectral index $n_\mathrm{s}$. We can achieve a similar effect by imposing a CMB prior on $n_\mathrm{s}$ \citep[analogous to][]{PhilcoxBOSS, PhilcoxFS+BAO2020}, essentially bringing both parameters closer to their fiducial values. Note that, in our case, degeneracies between $n_\mathrm{s}$ and other cosmological parameters are not as strong as in analyses of the BOSS data, since the largest scale we consider is $k_\mathrm{min} = 0.004$ $h$ Mpc$^{-1}$, while for BOSS $k_\mathrm{min} = 0.01$ $h$ Mpc$^{-1}$. In other words, due to the size of the simulation, we have more information from large-scale modes and can constrain the primordial slope better. We then impose a Planck prior on the scalar amplitude: separately as well as in combination with $n_\mathrm{s}$. The conclusion is that the unbiased cosmological parameters are reproduced only with the informative prior on the primordial parameter $A_\mathrm{s}$. Again, notice that the rest of nuisance parameters is only slightly affected by the choice of priors on $A_\mathrm{s}$ and $n_\mathrm{s}$: they remain unbiased and are not dominated by priors.

 When varying all cosmological parameters, we expect the inclusion of the bispectrum monopole will tighten the constraints on $A_\mathrm{s}$, $n_\mathrm{s}$, $A$, $\Omega_\mathrm{m} h^2$, $h$, because a) all of them are degenerate with each other, and improvement on one of them shrinks the contours for the others, b) the bispectrum significantly improves constraints on $b_1$ and alleviates the $w+A+A_\mathrm{s}+b_1$ degeneracy coming from the amplitude of the power spectrum, c) in contrast to the previous analyses in the literature, we include information from smaller scales, since our model is still valid up to $k_{\mathrm{max}, B}^{l=0} = 0.111$ $h$ Mpc$^{-1}$, hence the improvement of constraints on cosmological parameters in the case of our joint analysis will be more prominent than in the previous studies \citep[e.g., $\sim (5-15)\%$ improvement relative to the power spectrum analysis after the addition of the bispectrum monopole with  $k_{\mathrm{max}, B}^{l=0} = 0.08$ $h$ Mpc$^{-1}$ in][]{EFTIvanovBispectrum}. Additionally, we expect the bispectrum monopole to reduce the role of the choice of priors and decrease the effects of marginalisation by making the probability distribution closer to a multivariate Gaussian. 
A proper investigation of these effects is the topic of our future work.

\section{Conclusions}
\label{sec:conclusion}

In this paper we have performed a detailed analysis of the role of the bispectrum multipoles in constraining an interacting dark energy model. This is relevant for Stage-IV spectroscopic surveys, as a validation test and a comprehensive case study for the joint power spectrum and bispectrum analyses of extended cosmological models. 

We modelled the power spectrum and bispectrum multipoles with the EFTofLSS at one-loop and tree-level, respectively. We computed the linear growth factor and the logarithmic growth rate by solving the growth equation for the IDE model. As input data (the fiducial data vector in our likelihood pipeline) we used a large set of simulations at a single redshift $z=1$, complemented with a rescaled numerical covariance constructed from $10,000$ mock catalogues. Our fits used an effective volume that mimics the error budget of Stage-IV spectroscopic galaxy surveys, and we run our MCMC analyses with a Gaussian likelihood function. 

Using this setup, we first added more observables in sequence and studied them in terms of their ability to constrain the IDE parameters $w$ and $A$ without introducing a bias in the dark energy and nuisance parameters of the model. For the latter, we adopted as fiducial values the best fit values from previous studies performed in the context of standard $\Lambda$CDM \citep{MinervaBispectrumReal, MinervaPSBReal, MinervaBispectrumRSD, Chiara}. We found that the tree-level model for the bispectrum monopole is still valid up to $k \simeq 0.11~h~{\rm Mpc}^{-1}$, close to the largest value available in our measurements. In contrast to adding the hexadecapole to the lower-order multipoles of the power spectrum, the addition of the bispectrum monopole improved the constraints on the IDE parameters by $\sim 30\%$ for the equation of state parameter $w$, and $\sim 26\%$ for the coupling parameter $A$. We also found that the $P_0+P_2+P_4+B_0$ combination yields better constraints and less bias in the inferred parameters than $P_0+P_2+B_0$. 

In general, the large number of triangular configurations over which the bispectrum model has to be evaluated when including such nonlinear scales results in a rather computationally expensive analysis. Hence, we studied two approaches, which could allow us to get the same level of constraints but at smaller scales. The first approach aims at reducing the parameter space, which in our base case included 13 parameters, by means of bias relations. We tested the validity of two bias relations, and found that the tidal bias relation and its combination with the $b_2$ relation are the most suitable for this goal. In the second approach we added the bispectrum quadrupole to our Bayesian analysis. We found that either applying the bias relations or including the bispectrum quadrupole allows us to efficiently constrain the IDE parameter at more moderate values of $k_\mathrm{max}$ than in the base model $P_0+P_2+P_4+B_0$. For instance, to achieve the same level of constraints with FoM $\sim 6$ one can use either of the following: a) use only the bispectrum monopole in the base model, but include it up to high Fourier mode of $k_{\mathrm{max}, B}^{l=0} = 0.111$ $h$ Mpc$^{-1}$, b) apply the tidal bias relation in the scale range with the highest mode $k_{\mathrm{max}, B}^{l=0} = 0.095$ $h$ Mpc$^{-1}$, c) add the quadrupole measurements and evaluate the model up to $k_{\mathrm{max}, B}^{l=0, 2} = 0.09$ $h$ Mpc$^{-1}$.

We then forecasted the constraining power of Stage-IV spectroscopic galaxy surveys on the IDE parameters and the dark energy equation of state parameter in $w$CDM for fixed cosmological parameters. At $z=1$ we found $\sigma_w = 0.08$ and $\sigma_A = 2.51$ b GeV$^{-1}$ for the IDE case, while for the $w$CDM case, we found $\sigma_w = 0.1$. The first values are in a good agreement with the results of \citet{PedroCola} for the same redshift, where $\sigma_w = 0.06$ and $\sigma_A = 2$ b GeV$^{-1}$ were found when only power spectrum multipoles were used in the MCMC analysis. However, in \citet{PedroCola} the simulated data contained more than one order of magnitude less shot noise in comparison to our case, while the volume of the measurements was smaller with $V \sim 4$ Gpc$^3$ $h^{-3}$. Additionally, the covariance in the previous work was computed analytically up to the linear order, while the one used in this work is based on mocks, which is more realistic and contributes to larger uncertainties on the inferred parameters.

For comparison, we cite here the results of
the full-shape analyses (FS) with the EFTofLSS  from the BOSS survey of \citet{GuidowCDM} and \citet{Carrilho:2022}. We stress that the analysis of BOSS data takes advantage of two redshift bins and four different sky cuts \citep[three cuts in][]{GuidowCDM}, with the addition of BAO measurements, while the total volume of $6.5$ Gpc$^3$ $h^{-3}$ and mean number density of $2.2 \times 10^{-4}$ $h^3$ Mpc$^{-3}$ are very comparable with our values. In \citet{GuidowCDM}, the authors mention that FS analysis with the power spectrum alone is not sufficient to constraint $w$, due to strong degeneracies. They measure $w = -1.101^{+0.14}_{-0.11}$, which is still a stronger limit than the one obtained from Planck2018 data with lensing  \citep{Planck2018} alone: $w = -1.57^{+0.16}_{-0.33}$. These values of $w$ from the power spectrum analysis are in agreement with our forecast with fixed cosmological parameters without inclusion of the Alcock-Paczinsky effect, when the bispectrum is not included (see \autoref{fig:wCDM_full}). They also agree with the recent findings of \citet{Carrilho:2022} with $w = -1.17^{+0.12}_{-0.11}$ for $w$CDM without a CMB prior in the FS+BAO analysis. For IDE the authors of \citet{Carrilho:2022} find it necessary to include the CMB prior  on $A_\mathrm{s}$ and $n_\mathrm{s}$ (without BAOs), their analysis produces the following constraints: $w = -0.954^{+0.024}_{-0.046}$ and $A=4.8^{+2.8}_{-3.8}$ b GeV$^{-1}$ in the base model. In our setup for the analysis with power spectrum multipoles we find (see \autoref{fig:wACDM_AP_full}) $w = -1.14 \pm 0.09$ and $A = -4.45 \pm 3.9$ b GeV$^{-1}$. The discrepancy in the error on $w$ can be explained by the fact that the base model in \citet{Carrilho:2022} assumes informative priors on the nuisance parameters, when this assumption is alleviated and the priors are broader by 3 times (10 times) they find $w = -0.985^{+0.081}_{-0.038}$ ($w = -0.994^{+0.083}_{-0.046}$), which is closer to our uncertainties, despite them using BAO data in Table 8. In this work we assume broad flat priors on the nuisance parameters, hence our results are more in agreement with the BOSS analysis for uninformative priors on the nuisance parameters.

Without a CMB prior on the primordial parameters the model fails to accurately reproduce the true values used in the simulations in the case of the dark scattering model, while we do not observe similar behaviour in the case of the standard cosmological model or $w$CDM. The reason for that is a strong degeneracy between $A_\mathrm{s}$ and IDE parameters that drives the model prediction towards negative values of $w$ and $A$, small values in the scalar amplitude $A_\mathrm{s}$, and large values for the linear bias $b_1$. This hints to the presence of projection effects due to marginalisation of a strongly non-Gaussian posterior distribution. We demonstrate with synthetic data that such projection effect disappears when we include the bispectrum monopole to the analysis. Unfortunately, this is less prominent in the analysis with simulated data, since they already prefer more negative values of $w$, and hence of $A$. The fragile symmetry around $A=0$ and $w=-1$ is restored if the amplitude $A_\mathrm{s}$ is fixed to the fiducial value with an informative CMB prior. However, even in this case the addition of the bispectrum improves the constraints on $A$. This suggests that a joint power spectrum and bispectrum analysis can lead to improved constraints also on the other cosmological parameters, even when adopting informative priors on the primordial parameters. This will be the next step in our ongoing investigation.

A number of interesting questions remain to be answered. The first natural step would be to perform this analysis with the data from the BOSS survey to improve the constraints from the power spectrum analysis of \citet{Carrilho:2022} with the bispectrum. 
Additionally, it would be interesting to study the time-dependent dark energy equation of state, for example, in the CPL parametrisation \citep{w0waLinder, w0waPolarski}, which would allow for different effects of the interaction. This study requires data at different redshift bins, in order to constrain $w_0$ and $w_a$. In the case of IDE, the time-dependent equation of state promises to have weaker impact on the nonlinear scales than the constant-$w$ models \citep{IDESimulations2017}. Another potential direction of our study is the inclusion of one-loop effects in the bispectrum model, as was recently done for $\Lambda$CDM cosmology in \citet{EFTBispectrumOneLoop} and \citet{PhilcoxBispectrumOneLoop}. In the latter the authors show that the one-loop corrections for the bispectrum start to be significant already at $k_{\mathrm{max}, B}^{l=0}>0.1$ $h$ Mpc$^{-1}$. This is also expected to extend the reach of the model and therefore its constraining power.

\section*{Acknowledgements}

We are grateful to Claudio Dalla Vecchia and Ariel S\'{a}nchez for running and making available the Minerva simulations, performed on the Hydra and Euclid clusters at the Max Planck Computing and Data Facility in Garching, and to Pierluigi Monaco, for producing the Pinocchio mocks, run on the GALILEO cluster at CINECA thanks to an agreement with the University of Trieste. We thank Andrea Oddo and Emiliano Sefusatti for the effort in building the initial likelihood code, and for useful conversations. We further thank the referee for their useful comments and recommendation of a discussion on the degeneracies with cosmological parameters. We acknowledge use of the Cuillin computing cluster, Royal Observatory, University of Edinburgh.
We acknowledge use of open source software:
\texttt{Python} \citep{van1995python,Hunter:2007},  \texttt{numpy} \citep{numpy:2011},
\texttt{scipy} \citep{2020SciPy-NMeth}, 
\texttt{astropy} \citep{astropy:2018}, \texttt{corner} \citep{corner}, 
\texttt{GetDist} \citep{Lewis:2019xzd}. 
MT's research is supported by a doctoral studentship in the School of Physics and Astronomy, University of Edinburgh. 
AP is a UK Research and Innovation Future Leaders Fellow [grant MR/S016066/1]. PC and CM's research is supported by a UK Research and Innovation Future Leaders Fellowship [grant MR/S016066/1]. 
For the purpose of open access, the author has applied a Creative Commons Attribution (CC BY) licence to any Author Accepted Manuscript version arising from this submission.

\section*{Data Availability}

The data underlying this article will be shared on reasonable request to the corresponding author.

\bibliographystyle{mnras}
\bibliography{refs}

\appendix
\section{Full posterior distributions}\label{sec:app_full}

We collect here the full posterior distributions for the different runs described in the text.

\begin{figure*}
	\centering
	\includegraphics[width=\textwidth]{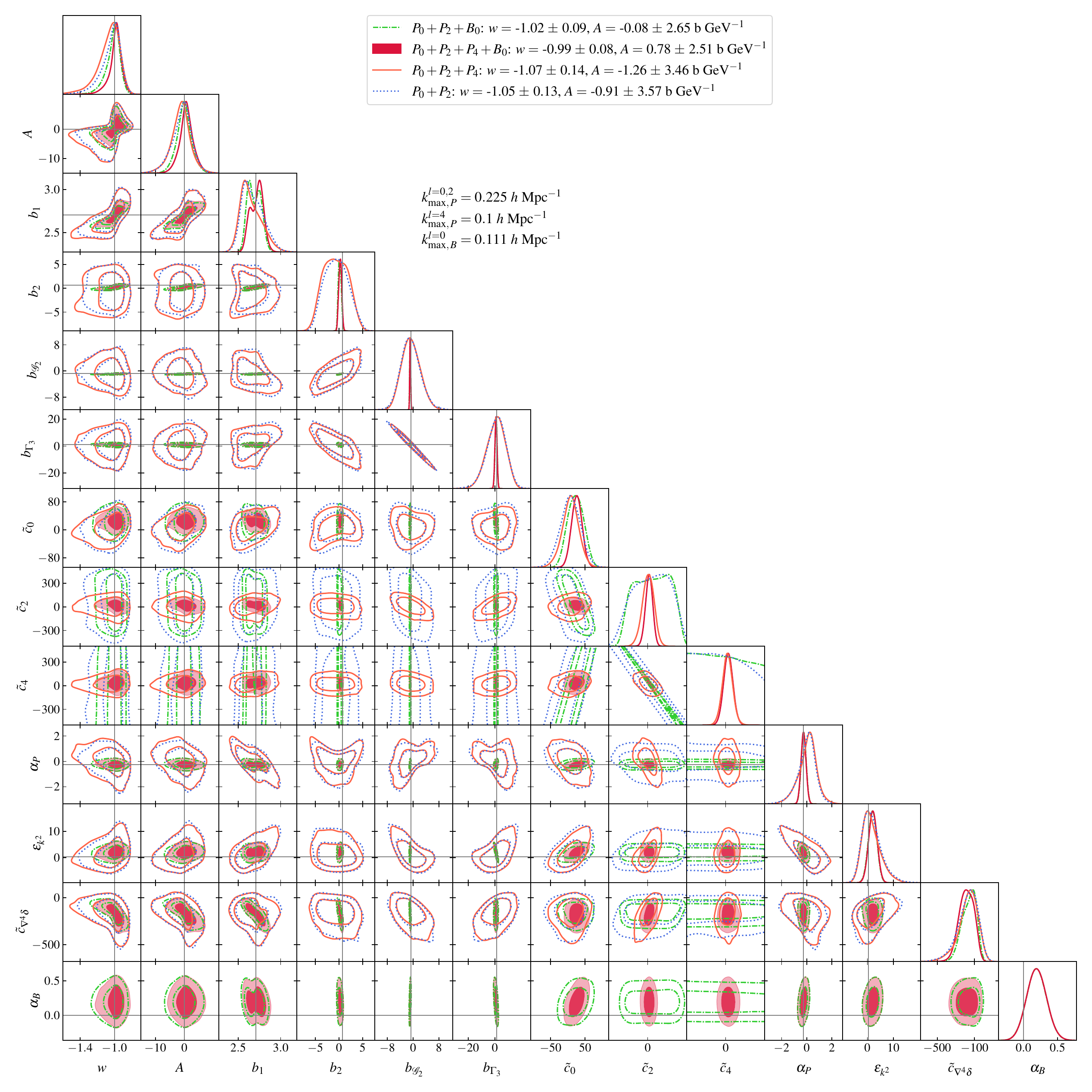}
    \caption{Posterior distributions for the base model with the scale cuts given in the triangle plot. The power spectrum monopole and quadrupole analysis is denoted by the dotted blue line, constraints from all power spectrum multipoles are given by the solid orange line, joint analysis of power spectrum monopole plus quadrupole and bispectrum monopole is presented by the dotted-dashed light-green line, and the full joint analysis is shown by the dark-red contours. Thin grey lines correspond to the fiducial values known from the fiducial cosmology and previous analysis of this data set in \citet{MinervaBispectrumReal, MinervaPSBReal, MinervaBispectrumRSD} .}
    \label{fig:base_full}
\end{figure*}

\begin{figure*}
	\centering
	\includegraphics[width=\textwidth]{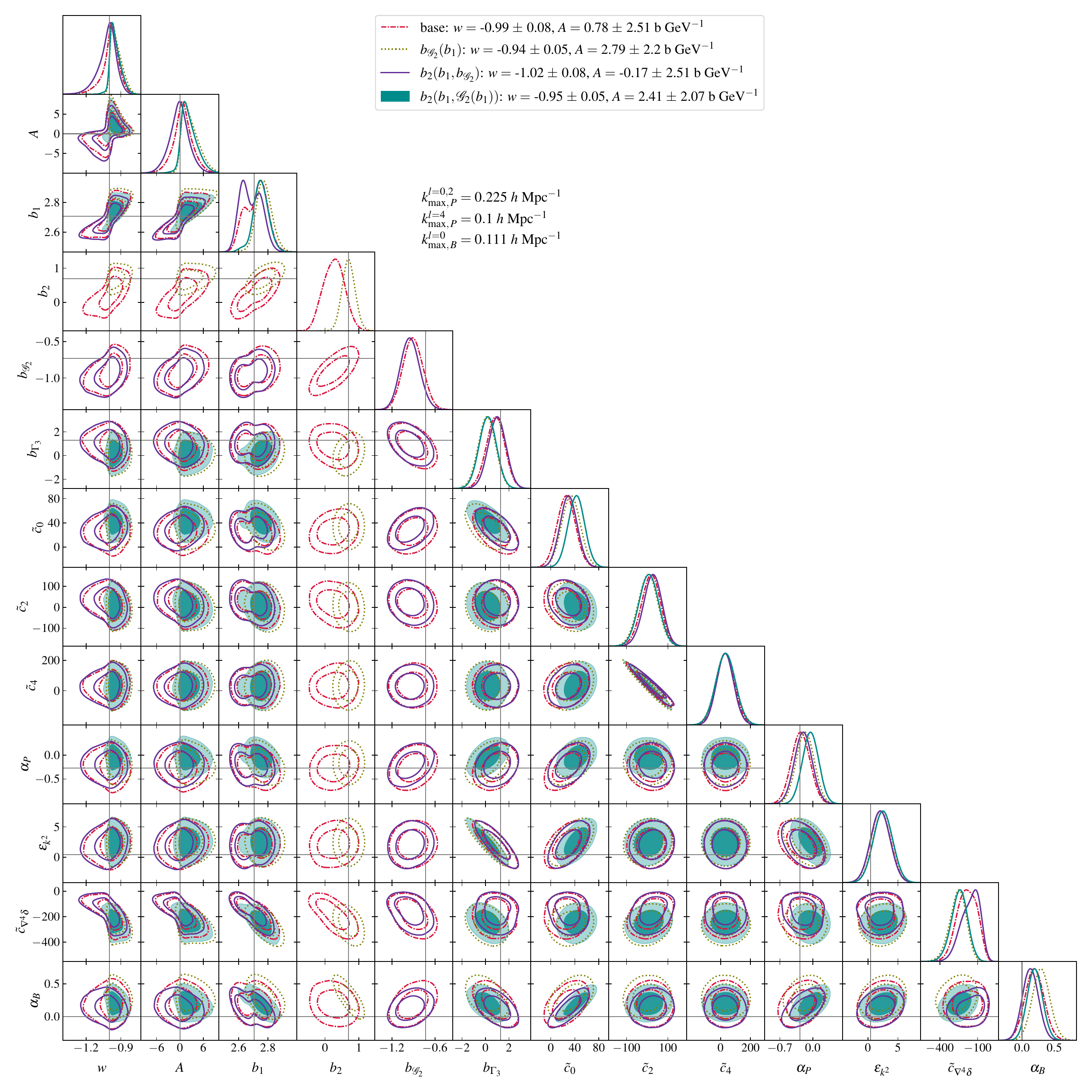}
    \caption{Posterior distributions for the $P_0+P_2+P_4+B_0$ models with bias relations at the scale cuts given on the triangle plot. The joint analysis with the base model is shown by the dark-red dotted-dashed lines, olive dotted lines denote the tidal bias relation, purple solid lines represent the $b_2$-relation, the dark-cyan contours denote the combined relation of $b_2$ and $b_{\mathcal{G}_2}$. Thin grey lines correspond to the fiducial values known from the fiducial cosmology and previous analysis of this data set in \citet{MinervaBispectrumReal, MinervaPSBReal, MinervaBispectrumRSD}.}
    \label{fig:biasrel_full}
\end{figure*}

\begin{figure*}
	\centering
	\includegraphics[width=\textwidth]{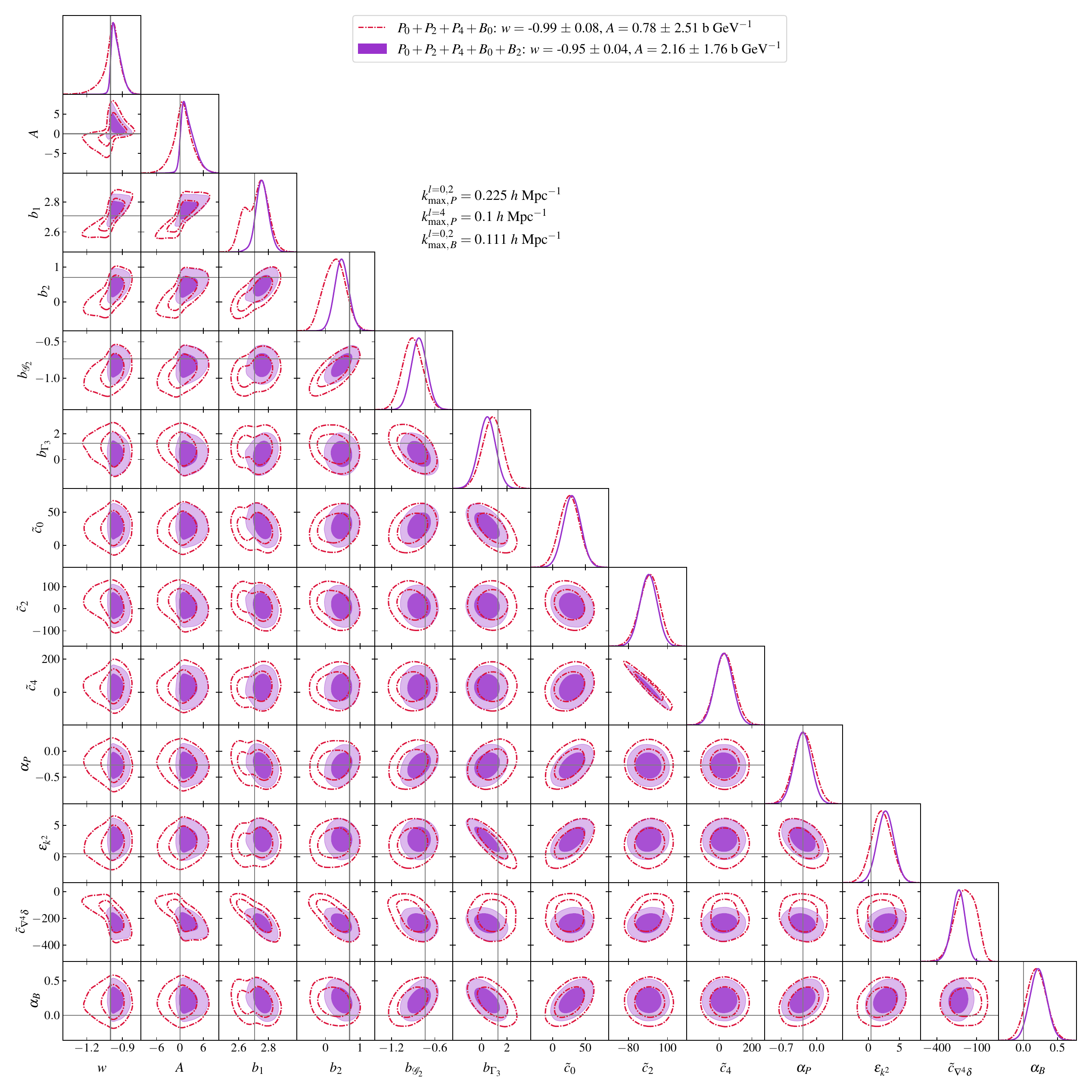}
    \caption{Posterior distributions for the base model with and without the bispectrum quadrupole at the scale cuts given on the triangle plot. The joint analysis with the bispectrum monopole is shown by the dark-red dotted-dashed lines, the joint analysis with the bispectrum monopole and quadrupole is denoted by the purple contours. Thin grey lines correspond to the fiducial values known from the fiducial cosmology and previous analysis of this data set in \citet{MinervaBispectrumReal, MinervaPSBReal, MinervaBispectrumRSD}.}
    \label{fig:B0B2_full}
\end{figure*}

\begin{figure*}
	\centering
	\includegraphics[width=\textwidth]{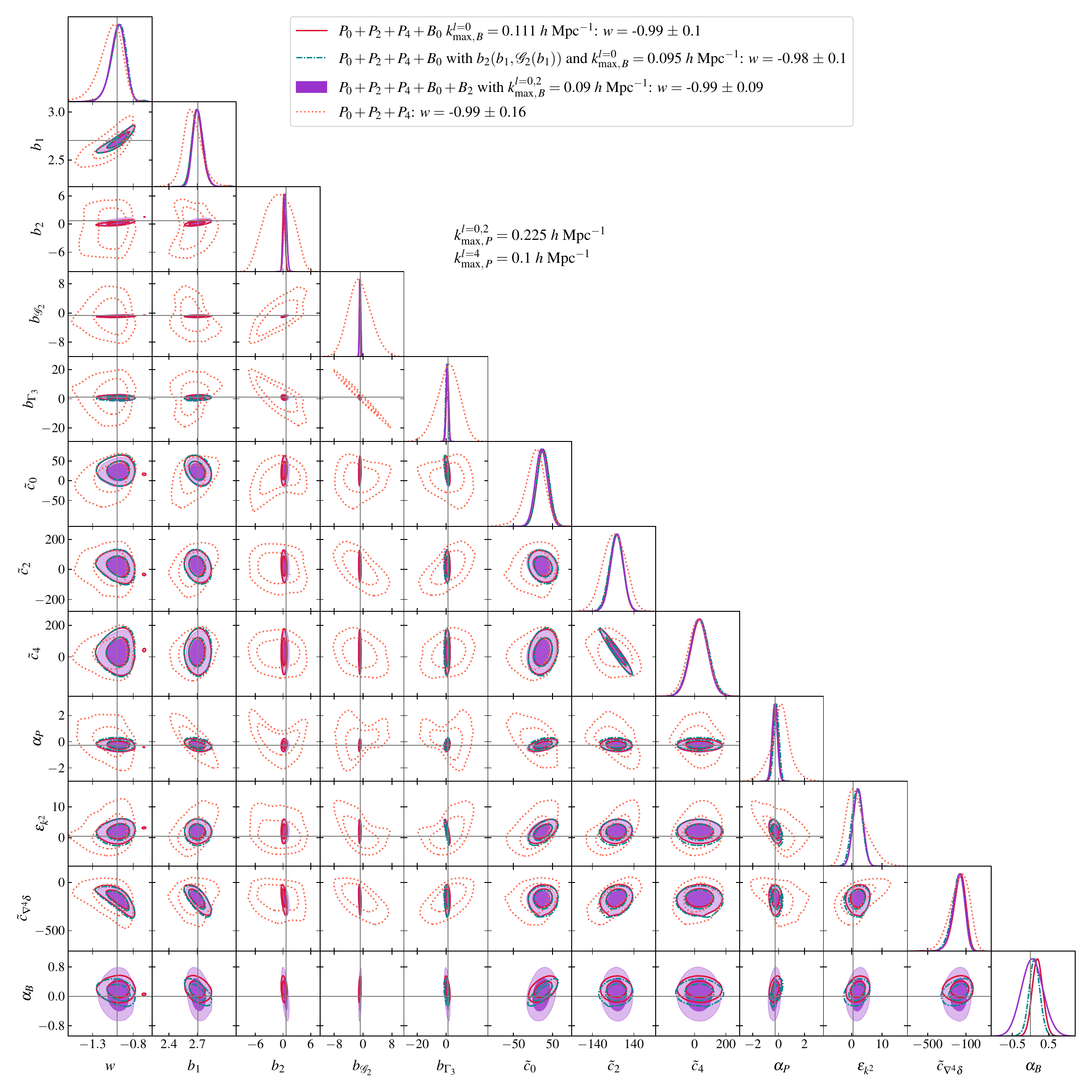}
    \caption{Posterior distributions for the $w$CDM model for four models at the scale cuts given on the triangle plot. The joint analysis with the bispectrum monopole is shown by the solid dark-red lines, the joint analysis with the bispectrum monopole with the combined bias relation is given by the dotted-dashed dark-cyan lines, the joint analysis with the bispectrum monopole and quadrupole is denoted by the purple contours, constraints from all power spectrum multipoles are shown by the dotted orange line. Thin grey lines correspond to the fiducial values known from the fiducial cosmology and previous analysis of this data set in \citet{MinervaBispectrumReal, MinervaPSBReal, MinervaBispectrumRSD}.}
    \label{fig:wCDM_full}
\end{figure*}

\begin{figure*}
	\centering
	\includegraphics[width=\textwidth]{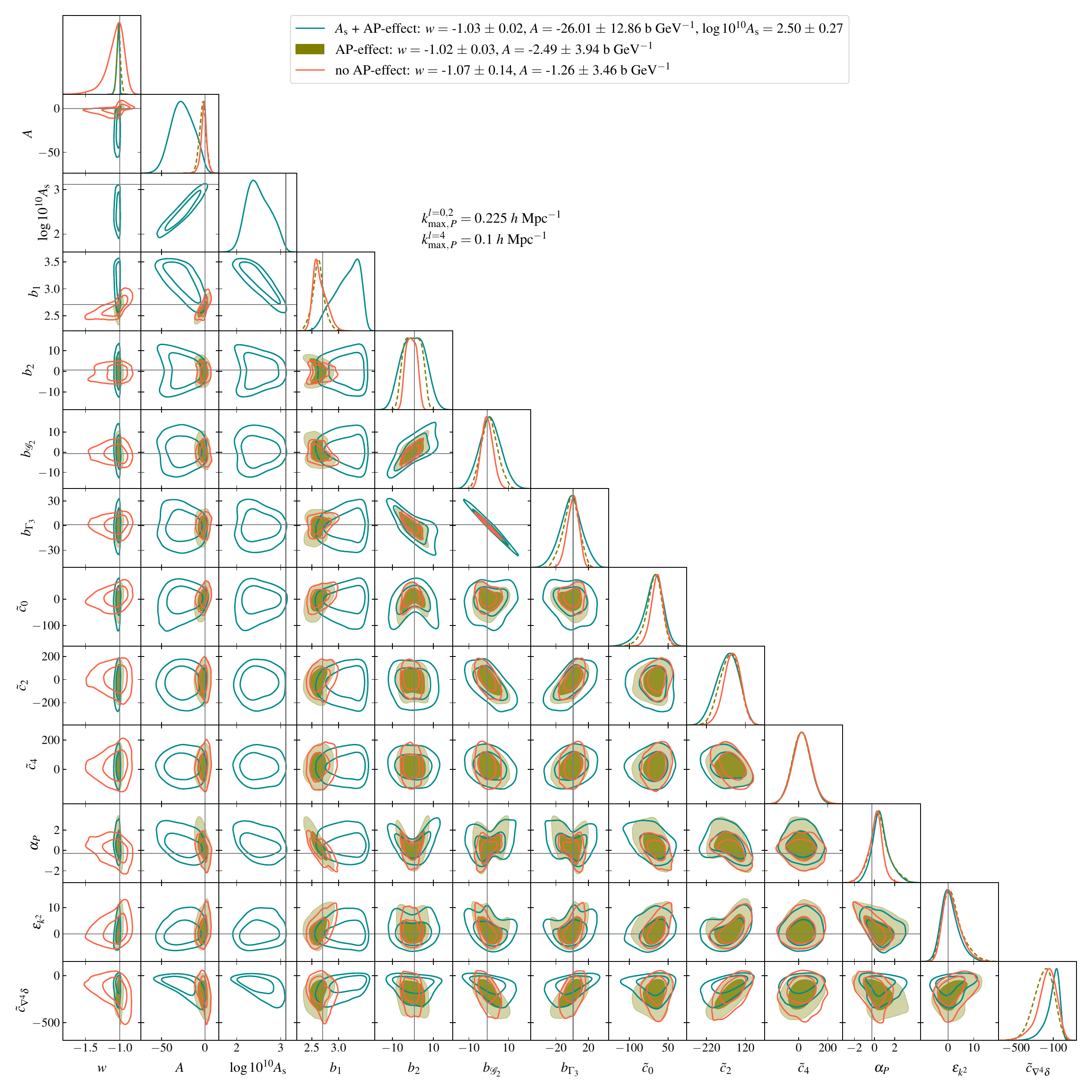}
    \caption{Posterior distributions for the demonstration of the AP-effect with the three power spectrum multipoles. The analysis with the variation of the scalar amplitude with a flat prior and addition of the AP-effect is shown by the solid dark-cyan lines, the analysis for variation of the IDE and nuisance parameters only with the addition of the AP-effect is shown with olive contours and constraints from all power spectrum multipoles without the AP-effect are given by the solid orange line (same as in \autoref{fig:base_full}). Thin grey lines correspond to the fiducial values known from the fiducial cosmology and previous analysis of this data set in \citet{MinervaBispectrumReal, MinervaPSBReal, MinervaBispectrumRSD}.}
    \label{fig:AP_full}
\end{figure*}

\begin{figure*}
	\centering
	\includegraphics[width=\textwidth]{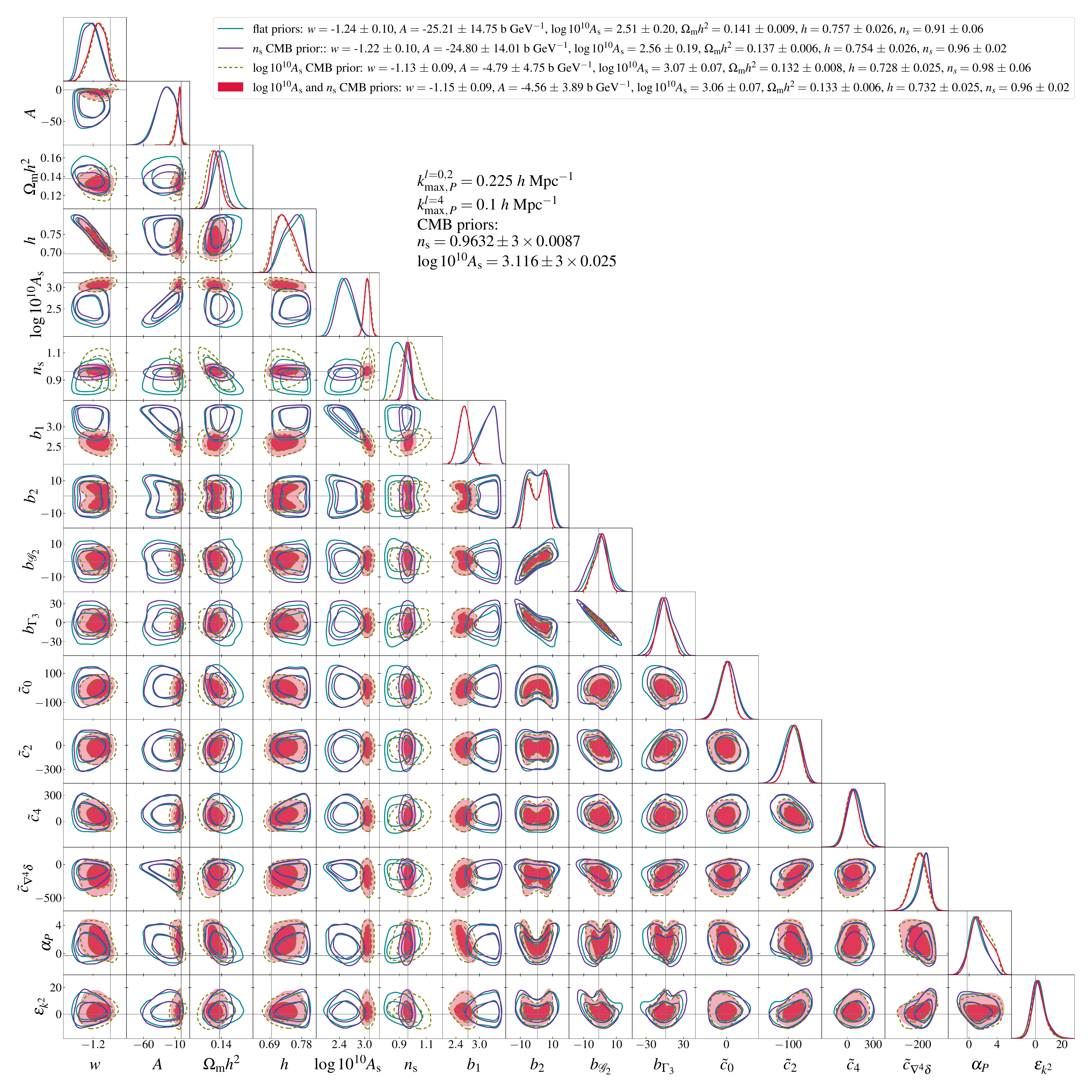}
    \caption{Posterior distributions for IDE model with the power spectrum multipoles only (the AP-effect included). The analysis with flat priors on all parameters is shown by the solid dark cyan lines, the analysis with the CMB prior on the primordial slope $n_\mathrm{s}$ is shown by the dark purple lines, the analysis with the CMB prior on the scalar amplitude $A_\mathrm{s}$ is shown by the olive dashed lines, the analysis with the CMB priors on both primordial parameters is shown by dark red contours. Thin grey lines correspond to the fiducial values known from the fiducial cosmology and previous analysis of this data set in \citet{MinervaBispectrumReal, MinervaPSBReal, MinervaBispectrumRSD}.}
    \label{fig:wACDM_AP_full}
\end{figure*}

\bsp
\label{lastpage}
\end{document}